\documentclass[twocolumn,showpacs]{revtex4}

\usepackage{latexsym}
\usepackage[dvips]{graphicx}
\usepackage{bm}
\usepackage{dcolumn}
\usepackage{bm}

\begin{document}
\hsize\textwidth\columnwidth\hsize\csname
@twocolumnfalse\endcsname

\title{ Noise-induced failures of chaos stabilization:
large fluctuations and their control}

\author{ I.A.~Khovanov$^{ab}$, N.A.~Khovanova$^{ab}$
 and P.V.E.~McClintock$^{b}$}

 \affiliation{$^a$ Physics Department, Saratov State University,
410026, Saratov, Russia\\
$^b$  Department of Physics, Lancaster University, LA1 4YB, UK}

\date{\today}

\widetext

\begin{abstract}

Noise-induced failures in the stabilization of an unstable orbit
in the one-dimensional logistic map are considered as large
fluctuations from a stable state. The properties of the large
fluctuations are examined by determination and analysis of the
optimal path and the optimal fluctuational force corresponding to
the stabilization failure. The problem of controlling
noise-induced large fluctuations is discussed, and methods of
control have been developed.

\end{abstract}

 \maketitle

\maketitle

\section*{Introduction}

The control of chaos represents a very real and important problem
in a wide variety of applications, ranging from neuron assemblies
to lasers and hydrodynamic systems \cite{Boccalettia:00}. The
procedure used consists of stabilizing an unstable periodic orbit
by the application of precisely designed small perturbations to a
parameter and/or a trajectory of the chaotic system. Different
methods of chaos control have been suggested and applied in many
different physical contexts, as well as numerically to model
systems \cite{Boccalettia:00}. For practical applications of these
control methods, it is important to understand how noise
influences the stabilization process, because fluctuations are
inherent and inevitably present in dissipative systems. The
problem has not been well studied. Typically, a method is
developed for stabilization of the orbit without initially taking
any account of fluctuations. Only then do the authors check the
robustness of their method by introducing weak noise into the
system \cite{Boccalettia:00}. Thus, in the celebrated pioneering
work of Ott, Grebogi and York, ``Controlling chaos" \cite{Ott:90},
the authors just noted that noise can induce failures of
stabilization.

In several works \cite{Bishop:96,Botina:97} methods are developed
for the stabilization of unstable orbits in the presence of noise.
They are based on a strong feedback approach to suppress any
deviation from the stabilized states. There are also methods
\cite{Collins:95} that use noise to move the system to a desired
unstable state, and then stabilize it there.

In this work we consider noise-induced failures in the
stabilization of an unstable orbit and the problem of controlling
these failures. The method of Ott, Grebogi and Yorke (OGY)
\cite{Ott:90} and a modification of the adaptive method (ADP)
\cite{Boccalettia:00} are used to stabilize an unstable point of
the logistic map. We consider the small noise limit where
stabilization failures are very rare and can therefore they be
considered as large fluctuations (deviations) from a stable state.
We study the properties of large deviations by determining the
optimal paths and the optimal fluctuational forces corresponding
to the failures. We employ two methods to determine the optimal
paths and forces. The first of these builds and analyzes the
prehistory probability distribution to determine the optimal path
and optimal force \cite{Dykman:92}. The second method considers an
extended map (relative to the initial one) which defines
fluctuational paths and forces in the zero-noise limit
\cite{Graham:91,Grassberger:89}. Furthermore we use the optimal
paths and forces to develop methods of controlling the large
deviations, i.e.\ the noise-induced failures of
stabilization\footnote{In the literature, methods for
stabilization are often referred to as a control methods too. To
differentiate controlling large fluctuations from controlling
chaos, we therefore use the term ``stabilization" to indicate the
control of chaos.}.

In section I we describe the procedures for local and global
stabilization of an unstable orbit of the logistic map. The
general approach to the control of a large deviation is presented
in section II. Noise-induced failures of local and global
stabilization are considered in sections III and IV respectively.
The results obtained are discussed in the conclusion.

\section{Chaos stabilization}

For simplicity we will stabilize an unstable fixed point $x^*$ of
the logistic map:
\begin{equation}
\label{lm}
x_{n+1}=r x_n(1-x_n),
\end{equation}
where $x_n$ is a coordinate, $n$ is discrete time and $r$ is the
control parameter that determines different regimes of the map's
behavior (\ref{lm}). The coordinate of the fixed point $x^*$ is
defined by the condition: $x_{n+1}=x_n$, and consequently its
location depends on the parameter $r$:
\begin{equation}
\label{cup}
x^*=1-\frac{1}{r}.
\end{equation}
We set the parameter $r=3.8$, a value for which an aperiodic
(chaotic) regime is observed (\ref{lm}), and the point $x^*$ is
embedded in the chaotic attractor.

From the range of existing stabilization methods, we chose to work
with just two: the OGY and ADP methods mentioned above.

To stabilize a fixed point by  the OGY method, perturbations
$\Delta r$ are applied to the parameter $r$, leading to the map
being modified (\ref{lm}) in the following manner:
\begin{eqnarray} \label{ogy} x_{n+1}&=&(r+\Delta r_n) x_n(1-x_n), \\ &&
\Delta r_n= r \frac{(2 x^* - 1) (x_n-x^*)}{x^* (1-x^*)}.\nonumber
\end{eqnarray}

To stabilize a fixed point by the ADP method, perturbations
$\Delta x$ are applied to the map's coordinate. The value of the
perturbation $\Delta x$ is defined by the distance between the
current system coordinate and the coordinate of the stabilized
state:
\begin{eqnarray}
\label{adp}
x_{n+1}&=&r x_n(1-x_n)+\Delta x_n, \\
&& \Delta x_n= (x_n-x^*).\nonumber
\end{eqnarray}
The ADP method is simple to use in practice. Different
modifications of the adaptive method are therefore used in many
papers devoted to experiments on the control of chaos.

We consider two types of stabilization procedure: local
stabilization and global stabilization.

During local stabilization, the perturbations $\Delta r$ and
$\Delta x$ differ from zero only if the following condition is
satisfied:
\begin{equation}
\label{loc} |x_n-x^*|< \epsilon.
\end{equation}
Here  $\epsilon$ is a small value: we fixed
$\epsilon=0.01$. If the condition (\ref{loc}) is not satisfied
then stabilization is absent, i.e. $\Delta r=0$ or $\Delta x=0$.

During global stabilization  perturbations are switched on when
the condition (\ref{loc}) is satisfied for the first time, and
remain present for all future time.

So, local or global stabilization involve modifications of the
initial map (\ref{lm}), and thus use another map in the form
(\ref{ogy}) or (\ref{adp}). The fixed point $x^*$ is an attractor
of the new map. After the stabilization is switched on, a
trajectory of the map tends to the fixed point $x^*$, and
subsequently remains there.

In the presence of noise the trajectory fluctuates in the vicinity
of the stabilized state, i.e.\ noise-induced dynamics appears. In
addition, noise can induce stabilization failures. For local
stabilization they imply a breakdown in the condition (\ref{loc}),
and for global stabilization they correspond to an escape of the
trajectory from the basin of attraction of the fixed point $x^*$.

Our aim is to study these noise-induced stabilization failures and
analyze the problem of how to suppress them. We therefore consider
the maps (\ref{ogy}) and (\ref{adp}) in the presence of additive
Gaussian fluctuations:
\begin{eqnarray}
\label{ogyn}
x_{n+1}&=&(r+\Delta r_n) x_n(1-x_n)+ D \xi_n, \\
&& \Delta r_n= r \frac{(2 x^* - 1) (x_n-x^*)}{x^* (1-x^*)},    \nonumber \\
\label{adpn}
x_{n+1}&=&r x_n(1-x_n)+\Delta x_n+ D \xi_n, \\
&& \Delta x_n= (x_n-x^*). \nonumber
\end{eqnarray}
Here $D$ is the noise intensity; and $\xi_n$ is a Gaussian random
process with zero-average  $\langle\xi\rangle=0$,
delta-correlation function
$\langle\xi_n\xi_{n+k}\rangle=\delta(k)$, and dispersion
$\langle\xi^2\rangle=1$. We use a high-speed noise generator
\cite{rt}.

\section{Control of large fluctuations}

Large fluctuations manifest themselves as large deviations from
the stable state of the system under the action of fluctuational
forces. Large fluctuations  play a key role in many phenomena,
ranging from mutations in DNA to failures of electrical devices.
In recent years significant progress has been achieved both in
understanding the physical nature of large fluctuations and in
developing approaches for describing them. The latter are based on
the concept of optimal paths -- the paths along which the system
moves during large fluctuations. Large fluctuations are very rare
events during which the system moves from the vicinity of a stable
state to a state remote from it, at a distance significantly
larger than the  amplitude of the noise. Such deviations can
correspond to a transition of the system to another state, or to
an excursion along some trajectory away from the stable state and
then back again. During such deviations the system is moved with
overwhelming probability along the optimal path under the action
of a specific (optimal) fluctuational force. The probability of
motion along any other (non-optimal) path is exponentially
smaller. In practice, therefore large fluctuations must of
necessity occur along deterministic trajectories. The problem of
controlling large fluctuations can thus be reduced to the task of
controlling motion along a deterministic trajectory. Consequently,
the control problem can be solved through application of the
control methods developed for deterministic systems
\cite{Pontryagin}.

Let us consider the control problem. Formally, the task that we
face in controlling noise-induced large fluctuations consists of
writing a functional $R$, the extrema of which correspond to
optimal solutions of the control problem, i.e.\ solutions with
minimal required energy
\cite{Whittle:96,Rabitz:97,Smelyanskiy:97}. The form of the
functional $R$ depends on a number of different additional
conditions related to e.g.\ the system dynamics, the energy of the
control force, or the time during which it is applied
\cite{Whittle:96,Rabitz:97,Smelyanskiy:97}. We will follow the
work \cite{Smelyanskiy:97} and consider the control of large
fluctuations by a weak additive deterministic control force.
Weakness means here that the energy of the control force is
comparable with the energy (dispersion) of the fluctuations (see
\cite{Whittle:96} for details). In this case, the extremal value
of the functional $R$ for optimal control, which moves the system
from an initial state $x^i$ to a target state  $x^f$, takes the
form \cite{Smelyanskiy:97}:
\begin{eqnarray}
\label{CF}
R_{\rm opt}(x^f,F)&=&S^{(0)}(x^f)\pm \Delta S, \\
\Delta S &=& (2 F)^{1/2}\left[ \sum_{k=N_i}^{N_f} (\xi_k^{\rm
opt})^2 \right]^{1/2}; \nonumber
\end{eqnarray}
where $\xi_k^{\rm opt}$ is the optimal fluctuational force that
induces the transition from $x^i$ to $x^f$ in the absence of the
control force; $S^{(0)}$ is an energy of the transition, $N_i$ and
$N_f$ are the times at which the fluctuational force $\xi_k^{\rm
opt}$ starts and stops \footnote{In our investigations we do not
assume any limitations on time moments $N_i$  and $N_f$.}, and $F$
is a parameter defining the energy of the control force.

The optimal control force $u_n^{\rm opt}$ for the given functional
(\ref{CF}) is defined \cite{Smelyanskiy:97} by:
\begin{eqnarray}
\label{OptF} u_n^{\rm opt}= &\mp& (2F)^{1/2} \xi_n^{\rm opt}\left[
\sum_{k=N_i}^{k=N_f} (\xi_k^{\rm opt})^2 \right]^{1/2} \nonumber \\
&\times& \delta(x_n-x_n^{\rm (0)opt}) \, ,
\end{eqnarray}
where $x_n^{\rm (0)opt}$ is the optimal fluctuational path in the
absence of the control force. The minus sign in the expression
(\ref{OptF}) decreases the probability of a transition to the
state $x^f$, and the plus sign increases the probability. It can
be seen (\ref{OptF}) that the optimal control force $u_n^{\rm
opt}$ is completely defined by the optimal fluctuational force
$\xi_k^{\rm opt}$, and the optimal fluctuational path $x_n^{\rm
(0)opt}$, corresponds to the large fluctuation. Therefore to solve
the control problem it is necessary, first, to determine the
optimal path $x_n^{\rm (0)opt}$ leading from the state $x^i$ to
the state $x^f$ under the action of the optimal fluctuational
force $\xi_k^{\rm opt}$. Thus, a solution of the control problem
depends on the existence of an optimal path: it is obvious that
the approach described should be straightforward to apply,
provided that the optimal path exists and is unique.

We consider below an application of the approach described to
suppress large fluctuations in the one-dimensional map. The large
fluctuations in question are considered here to correspond to
failures in the stabilization of an unstable orbit.

The control procedure depends on the determination of the optimal
path and optimal fluctuational force and, to define them, we will
use two different methods. The first is based on an analysis of
the prehistory probability distribution (PPD) and the second one
consists of solving a boundary problem for an extended map which
defines fluctuational trajectories.

The PPD was introduced in \cite{Dykman:92} to analyze optimal
paths experimentally in flow systems. We will use the distribution
to analyze fluctuational paths in maps. Note, that in
\cite{Luchinsky:97,Khovanov:00} it was shown that analysis of the
PPD allows one to determine both the optimal path and the optimal
fluctuational force. The essence of this first method consists of
a determination of the fluctuational trajectories corresponding to
large fluctuations for extremely small (but finite) noise
intensity, followed by a statistical analysis of the trajectories.
In this experimental method the behaviour of the dynamical
variables $x_n$ and of the random force $\xi_n$ are tracked
continuously until the system makes its transition from an initial
state $x^i$ to a small vicinity  of the target state $x^f$. Escape
trajectories $x_n^{\rm esc}$ reaching this state, and the
corresponding noise realizations $\xi_n^{\rm esc}$ of the same
duration, are then stored. The system  is then reset to the
initial state $x^i$ and the procedure is repeated. Thus, an
ensemble of trajectories is collected and then the fluctuational
PPD $p_n^h$ is constructed for the time interval during which the
system is monitored. This distribution contains all information
about the temporal evolution of the system immediately before the
trajectory arrives at the final state $x^f$. The existence of an
optimal escape path is diagnosed by the form of the PPD $p_n^h$:
if there is an optimal escape trajectory, then the distribution
$p_n^h$ at a given time $n$ has a sharp peak at optimal trajectory
$x_n^{\rm opt}$. Therefore, to find an optimal path it is
necessary to build the PPD and, for each moment of time $n$, to
check for the presence of a distinct narrow peak in the PPD. The
width of the peak defines the dispersion $\sigma_n^h$ of the
distribution and it has to be of the order of the mean-square
noise amplitude $\sqrt{D}$ \cite{Dykman:92}. The optimal
fluctuational force that moves the system trajectory along the
optimal path can be estimated by averaging the corresponding noise
realizations $\xi_n^{\rm esc}$ over the ensemble. Note, that
investigations of the fluctuational prehistory also allows us to
determine the range of system parameters for which optimal paths
exist.

To determine the optimal path and force by means of the second
method we analyze extended maps \cite{Graham:91,Grassberger:89}
using the principle of least action \cite{Grassberger:89}. Such
extended maps are analogous to the Hamilton-Jacobi equation in the
theory of large fluctuations for flow systems. For the
one-dimensional map $x_{n+1}=f(x_n)+D\xi_n$, the corresponding
extended map in the zero-noise limit takes the form:
\begin{eqnarray}
\label{extmap}
x_{n+1}&=&f(x_n)+y_n/g(x_n), \nonumber \\
y_{n+1}&=&y_n/g(x_n), \\
 g(x_n) & = &\frac{\partial f(x_n)}{\partial x_n}. \nonumber
\end{eqnarray}
The map is area-preserving, and it defines the dynamics of the
noise-free map $x_{n+1}=f(x_n)$, if $y_n=0$. If $y_n\ne 0$ then
the coordinate $x_n$ corresponds to a fluctuational path, and the
coordinate $y_n$ to a fluctuational force. Stable and unstable
states of the initial map become saddle states of the extended
map. So, the fixed point $x^*$ of the ADP (\ref{adpn}) and OGY
(\ref{ogyn}) maps becomes a saddle point of the corresponding
extended map. Fluctuational trajectories (including the optimal
one) starting from $x^*$ belong to unstable manifolds of the fixed
point $(x^*,0)$ of the extended map.

The procedure for determination of the optimal paths consists of
solving the boundary problem for the extended  map (\ref{extmap}):
\begin{eqnarray}
\label{bc1}
x_{-\infty}=x^*, ~~~~~  y_{-\infty}=0 \\
\label{bc2}
x_{\infty}=x^f, ~~~~y_{\infty}=0;
\end{eqnarray}
where $x^*$ is the initial state and $x^f$ is a target state.

To solve the boundary problem different methods can be used. For
the one-dimensional maps under consideration, a simple shooting
method is enough \cite{numeric}. We choose an initial perturbation
$l$ along the linearized unstable manifolds in a vicinity of the
point $(x^*,0)$ of the map (\ref{extmap}). The procedure to
determine a solution can be as follows: looking over all possible
values $l$, we determine a trajectory which tends to the point
$(x^f,0)$. Note that, because these maps are irreversible there
exist, in general, an infinite number of solutions of the boundary
problem. The optimal trajectory (path) has minimal action (energy)
$S=\sum_{n=-\infty}^{\infty}y_n^2$; here $y_n$ is calculated along
the trajectory, corresponding a solution of the boundary task.

\section{Noise-induced failures in local stabilization}

A breakdown of the condition (\ref{loc}) corresponds to a failure
of local stabilization, i.e.\ to the noise-induced escape of the
trajectory from an $\epsilon$-vicinity of the fixed point  $x^*$.
The target state $x^f$ corresponds to the boundaries of the
stabilization region: $x^f=x^*\pm\epsilon$.

Instead of analyzing the maps (\ref{ogyn}) and (\ref{adpn}) in the
$\epsilon$-vicinity of the fixed point $x^*$  we can
investigate linearized maps of the following form:
\begin{eqnarray}
\label{linn}
x_{n+1}&=&a x_n+ D \xi_n;
\end{eqnarray}
here $a$ is a value of derivative $\partial f(x_n)/\partial x_n$
in the fixed point $x^*$. For the map (\ref{ogyn}) the derivative
is equal to zero $a_{\rm OGY}=0$, and for the map (\ref{adpn})
$a_{\rm ADP}=-0.8$.

Let us investigate stabilization failure by considering the most
probable (optimal) fluctuational  paths, which lead from the point
$x^*$ to boundaries $x^*\pm \epsilon$. For linearized maps
(\ref{linn}) the extended map (\ref{extmap}) can be reduced to the
form:
\begin{eqnarray}
\label{aux}
x_{n+1}=ax_n+\frac{y_n}{a}, \\
y_{n+1}=\frac{y_n}{a} \nonumber
\end{eqnarray}
with the initial condition  ($x_0=x^*$, $y_0=0$) and the final
condition  $x^f=x^* \pm \epsilon$. It can be seen that a solution
of the map (\ref{aux}) increases proportionally to $y_n= {\rm
const}/a^n$ \cite{Reimann:91}. This means that, for the ADP map
(\ref{adpn}), the amplitude of the fluctuational force increases
slowly but that, for the OGY map (\ref{ogyn}), the failure arises
as the result of only one fluctuation (iteration). Because
equation (\ref{aux}) is linear, the boundary problem will have a
unique solution \cite{numeric}. Thus, analysis of the linearized
extended map (\ref{aux}) shows that there is an optimal path, and
it gives a qualitative picture of exit through the boundary
$x^*\pm \epsilon$.

Let us check the existence of the optimal paths through an
analysis of the prehistory of fluctuations. To obtain exit
trajectories and noise realizations we use the following
procedure. At the initial moment of time, a trajectory of the map
is located at point $x^*$. The subsequent behaviour of the
trajectory is monitored until the moment at which it exits from
the $\epsilon$-region of the point $x^*$. The relevant part of the
trajectory, just before and after its exit, are stored. The time
at which the exit occurs is set to zero. Thus ensembles of exit
trajectories and of the corresponding noise realizations are
collected and PPDs are built.

To start with, we will discuss these ideas in the context of the
ADP map. Fig.~1(a) shows PPDs of the escape trajectories of the
ADP map, and the corresponding noise realizations for the exit
through the boundary $(x^*-\epsilon)$ are shown in Fig.\ 1(b). The
picture of exit through the other boundary $(x^*+\epsilon)$ is
symmetrical, so we present results for one boundary only. It is
evident (Fig.~1) that there is the only one exit path. Note, that
the path to the boundary $(x^*-\epsilon)$ is approximately
2.8$\times$ more probable than the path to the boundary
$(x^*+\epsilon)$. This difference arises from an asymmetry of the
map in respect of the boundaries.

Because for each boundary there is the only one exit path, the
optimal path and the optimal fluctuational force can be determined
by simple averaging of escape trajectories and noise realizations
respectively. In Fig.~2 the optimal exit paths and the optimal
fluctuational forces are shown for the boundaries $(x^* -
\epsilon)$ and $(x^* + \epsilon)$. The paths and the forces
coincide with a solution of the boundary problem (circles in the
Fig.~2) of the extended linear map (\ref{aux}). The time
dependence of the dispersion $\sigma_n^h$ of  PPDs for the exit
trajectories and noise realizations are shown in Fig.~3. As can be
seen (Fig.~2) the optimal path is long,  and the amplitude of the
fluctuational force increases slowly, in agreement with analysis
of the linearized map (\ref{aux}). The dispersion $\sigma_n^h$ of
both trajectories and noise realizations decreases by construction
as the boundary is achieved (Fig.~3).

The optimal fluctuational force obtained (Fig.~2(b)) must
correspond \cite{Khovanov:00} to the energy-optimal deterministic
force that induced the stabilization failure. We have checked this
prediction and found that the optimal force induces the exit from
an $\epsilon$-region of the point $x^*$: we selected an initial
condition at the point $x^*$ and included the optimal
fluctuational force additively; as a result we observed the
stabilization failure. If we decrease the amplitude of the force
by 5-10\%, then the failure does not occur. It appears, therefore,
the deduced force allows us to induce the stabilization failure
with minimal energy (see  \cite{Khovanov:00} for details).

Using the optimal path and the force we can solve the opposite
task \cite{Whittle:96,Smelyanskiy:97} --- to decrease the
probability of the stabilization failures. Indeed, if during the
motion along the optimal path we will apply a control force with
the same amplitude but with the opposite sign as the optimal
fluctuational force has, then, obviously, the failure will not
occur. Because we know the optimal force then, in accordance with
the algorithm \cite{Smelyanskiy:97} described above, it is
necessary to determine the time moment when system is moving along
the optimal path. For the ADP method the optimal path is long
enough to identify that a trajectory is moving along the optimal
path, and then to apply a control force.

In the presence of control the map (\ref{adpn}) is modified:
\begin{eqnarray}
\label{adpncont}
x_{n+1}&=&r x_n(1-x_n)+\Delta x_n+ D \xi_n +u_n, \\
&& \Delta x_n= (x_n-x^*); \nonumber
\end{eqnarray}
here $u_n$ is the deterministic control force.

We use the following scheme to suppress the stabilization
failures. Initially the control force is equal to zero ($u_n=0$)
and the map is located in the point $x^*$; we continuously monitor
a trajectory of the map (\ref{adpncont}) and define the time
moment when the system starts motion along the optimal path
$\langle x_n \rangle$. We assume that the system moves along the
optimal path $\langle x_n \rangle$ if it passes within a small
vicinity of the coordinate $\langle x_{-2} \rangle$ and then
within a small vicinity of $\langle x_{-3} \rangle$ (see arrows in
Fig.~2(a)). Then on the following iteration we add the control
force $u_n=- {\rm sign}(\xi_n)\langle \xi_n \rangle$, $n=-1$ (see
Fig.~2(b)).

In Fig.~4(a) dependences of the mean time $\langle \tau\rangle $
between the failures on the noise intensity $D$ are plotted in the
absence, and in the presence, of the control procedure. It is
clear that the mean time $\langle \tau\rangle $ is substantially
increased by the addition of the control, i.e.\ stability in the
face of fluctuations is significantly improved by the addition of
the control scheme. The efficiency of the control procedure
depends exponentially \cite{Smelyanskiy:97} on the amplitude of
the control force (Fig.~4(b)), and there is an optimal value of
the control force, which is very close to the value (arrow in
Fig.~4(b)) of the optimal fluctuational force.

Now consider noise-induced stabilization failures for OGY map
(\ref{ogyn}). An analysis of the linearized map has shown that the
failure occurs as the result of a single fluctuation. We have
checked the conclusion by an analysis of the fluctuational
trajectories of the map (\ref{ogyn}), much as we did for the ADP
map. The optimal path and optimal force are shown in Fig.~5 for
both boundaries, $(x^*+\epsilon)$ and $(x^*-\epsilon)$. An exit
occurs during one iteration and there is no a prehistory before
this iteration.  It means that we cannot determine the moment at
which the large fluctuation starts and, consequently, that we
cannot control the stabilization failures. The existence of a long
prehistory is thus a key requirement in the control the large
fluctuations.

We can of course decrease the probability of a failure by
increasing the $\epsilon$-region of stabilization. The maximum
possible increase would correspond to infinite boundaries -- in
which case we would be dealing with global stabilization.

\section{Noise-induced failures of global stabilization}

To investigate fluctuational dynamics in the global stabilization
regime, we consider the dynamics of the maps (\ref{ogyn}) and
(\ref{adpn}) with initial conditions at the fixed point $x_0=x^*$.
We will first consider them in the absence of noise. The maps are
shown on the plane $x_n-x_{n+1}$ in the Fig.~6.

The map (\ref{ogyn}) (Fig.~6(a)) has three fixed points of
period one: the point $x^*\approx 0.7368$ is stable with the
multiplier $\mu =0$; the points $x^*_2=0$ and  $x^*_1 \approx
0.5906 $ are unstable with multipliers $\mu\approx -3.04$ and $\mu
\approx 1.8016$, respectively. The map has two attractors: the
point $x^*$ and the attractor at infinity \footnote{Reaching the
attractor at infinity can be viewed as a transition of the system
to another regime, not described by the mathematical model.}.
Their basins of attraction (Fig.6~(a)) are self-similar
(fractal) \cite{Kaneko:83,Takesue:84}. The point $x^*_1$ and its
pre-images by backward iteration lie on the basin boundaries of
the attractors \cite{Grebogi:87}. In the intervals $x \in
(-0.183,0.5906)$ and $x \in (0.862,1.027)$ the basins of the
attractors alternate and are of different length. The interval $x
\in (0.5906,0.862)$ corresponds to the widest basin of the fixed
point $x^*$. The boundaries of this basin are defined by the
unstable point $x^*_1$ and its pre-image $x^{I*}_1$. The
semi-infinite intervals $ x\in (-\infty,-0.183)$ and $x \in
(1.027,\infty) $ correspond to basins of the attractor at
infinity. The boundaries of the semi-infinite intervals are
defined by the points $x^{-\infty}= -0.183$ and
$x^{\infty}=1.027$, which correspond to the cycle of period 2.

The map  (\ref{adpn}) (Fig.~6(b)) has  two fixed points: the
point $x^*\approx 0.7368$ is stable with multiplier $\mu \approx
-0.8$; and the point $x^*_1 \approx 0.2632$ is unstable with
multiplier $\mu \approx 2.8$. The map has two attractors: the
fixed point $x^*$ and the attractor at infinity. The basins of
attraction are smooth (Fig.~6~(b)). The first boundary of the
basins is the point $x^*_1$ and the second boundary is a pre-image
$x^{I*}_1$ of the point $x^{*}_1$.

So, each of the maps has two attractors, but the structure of
their basins of attraction are qualitatively different.

We now consider these maps (\ref{ogyn}) and (\ref{adpn}) in the
presence of noise. Noise can induce escape from the basin of the
fixed point $x^*$, corresponding to failure of the stabilization.
As before we examine the dynamics of the escape trajectories
obtained for extremely small noise intensity in order to determine
the optimal path and the optimal force. Fluctuational escape
trajectories of the map (\ref{adpn}) are shown by dots on the
plane $(x_{n}-x_{n+1})$ in Fig.~6(b).  As can be seen, there is
one escape path, and the escape trajectories pass through the
unstable point $x^*_1$.  In Fig.~7 the optimal path and the
optimal force obtained by averaging the escape trajectories and
noise realizations respectively are shown by crosses. The
stabilization failure clearly possesses a long prehistory. From
the point of view of the control procedure, the presence of a
large deviation of the system coordinate $\langle x_n \rangle$ at
the time moment $n=-1$, and the smaller deviation of the
fluctuational force $\langle \xi_n \rangle$ at the next time
moment ($n=0$), are important.This is because the  first
fluctuation of coordinate $x_n$ can easily be identified and
distinguished from non-optimal fluctuations in the vicinity of the
stable state $x^*$.

Next, we examine the process of escape for the map (\ref{ogyn}).
Fig.~8(a) shows escape trajectories superposed at the time moment
when the trajectory crosses the basin boundary at the point
$x^{-\infty}$. It is evident that there is no selected escape
path. The escape trajectories can be divided into several groups
with different probabilities. With maximum probability (almost
50\%) the escape trajectories follow the arrowed path in
Fig.~8(a)) corresponding to motion in the direction of the point
$x^{-\infty}$ without any jumps in the opposite direction. The
other paths include jumps in the opposite direction. The width of
the distribution of fluctuational paths is comparable with the
noise amplitude and there is no a specific fluctuational force. In
Fig.~6(a) the escape trajectories are shown on the plane $(x_n$ --
$x_{n+1})$. Is can be seen that, after the point $x_1^*$, the
escape trajectories are located close to the trajectories of
deterministic map, so we can suppose that after the point $x_1^*$
the motion has the character of directed diffusion. The interval
between the points  $x_1^*$  and $x^{-\infty}$ lies within the
fractal basin, and this fact implies a variety of escape paths.
Indeed, within a small vicinity of the point $x_1^*$ there is a
piece of basin of the attractor at infinity. For escape,
therefore, it is enough to bring the trajectory only to this
basin. However, the size of this basin is small and a weak
fluctuation can of course move the trajectory back to the basin of
point $x^*$ and vice versa. As a result, the trajectory can spend
a long time in the vicinity of the point $x_1^*$: it can return to
the point $x^*$, as well as escape from the basin of the point
$x^*$.

Thus, the fractal structure in the basin of attraction leads to
complex behaviour of the escape trajectories; they can spend a
long time in the fractal basin; motion in the direction of the
attractor at infinity has the largest probability.

Investigations of escape from the point $x^*$ to the vicinity of
the point $x_1^*$ have shown (Fig.~8(b)), that there is no
specific path within this interval, so that we cannot determine
the optimal path or the optimal force using an analysis of the
escape trajectories. It is possible to select several different
favoured paths (thick lines in the Fig.~8(b)), but dispersion of
the trajectories for each of them is much larger than the noise
intensity used.

We now determine the escape optimal paths and the forces by
solving the boundary problems (\ref{bc1}) and (\ref{bc2}) for the
extended maps:
\begin{eqnarray}
\label{hamogy}
x_{n+1}&=&f(x_n)+y_n/g(x_n), \nonumber \\
y_{n+1}&=&y_n/g(x_n), \\
f(x_n)&=&(r+\Delta r_n) x_n(1-x_n), \nonumber \\
g(x_n)&=&\frac{\partial f(x_n)}{\partial x_n} \nonumber
\end{eqnarray}
and
\begin{eqnarray}
\label{hamadp}
x_{n+1}&=&f(x_n)+y_n/g(x_n), \nonumber \\
y_{n+1}&=&y_n/g(x_n), \\
f(x_n)&=&r x_n(1-x_n)+\Delta x_n, \nonumber \\
g(x_n)&=&\frac{\partial f(x_n)}{\partial x_n}, \nonumber
\end{eqnarray}
which correspond to the maps (\ref{ogyn}) and (\ref{adpn}).
In such a way we have used the extended map (\ref{aux}) to analyze the
linearized map (\ref{linn}).

First, we consider the results of solving the boundary problem for
the extended map (\ref{hamogy}). To do so, we use a shooting
method, with boundary conditions (\ref{bc1}) and (\ref{bc2}),
where $x^f=x^*_1$.  Since the derivative $g(x_n)=\frac{\partial
f(x_n)}{\partial x_n}$ of the map (\ref{ogyn}) at the point $x^*$
is equal to zero, we cannot calculate eigenvectors of the point
$(x^*,0)$ of the map (\ref{hamogy}).  Therefore, as a parameter of
the boundary problem we choose an initial perturbation $y_0$,
since it defines all the trajectories going away from the point
$(x^*,0)$. Four solutions of the boundary problem, obtained
numerically, are found to have practically the same action $S$.
Four escape paths and noise realizations ($t_1-t_4$) of the map
(\ref{hamogy}) corresponding to these solutions are shown in
Fig.~9. The trajectory $t_4$ has the minimum activation energy
$S\approx 0.0115$ and the energies of other trajectories are
practically the same: $S\approx 0.0123$. All the optimal
trajectories lie on a stable manifold of the point ($x_1^*,0$),
and the stable manifold goes to the point ($x^*,0$) (Fig.~10). If
we take into account the fact that the noise intensity is finite
during the experimentally analysis of escape trajectories
(Fig.~8), then the fluctuational trajectories of the map
(\ref{ogyn}) form a wide bunch around the optimal paths and
trajectories can go along different the optimal paths at different
time intervals. Thus, for the OGY map (\ref{ogyn}), the only way
to determine optimal paths and forces is by solution of the
boundary problem for the extended map, whereas an analysis of the
PPD is not successful.

Now, let us consider the solution of the boundary problem for the
map (\ref{hamadp}).  We have defined an unstable direction of the
point $(x^*,0)$ and used the length of a vector $l$ along this
direction as a parameter of the boundary problem. There is just
one solution for which the value of action $S=0.0449$, which is
slightly smaller than the value $S=0.493$ calculated by using the
PPD. The corresponding optimal path and optimal force are shown in
Fig.~7 together with the path and the force found by using PPD. It
can be seen (Fig.~7), that the optimal paths and the forces
obtained by calculated PPD and by using the extended map are
practically the same.

Thus, we have defined the optimal path and the optimal force
corresponding to global stabilization failures, and we have
compared two methods for determination of the optimal path and
force: the first method being based on an experimental analysis of
the prehistory probability distribution, and the second one being
based on solving the boundary problem for an extended
area-preserving map. The latter method allows us to determine the
optimal path and force for both the maps (\ref{ogyn}) and
(\ref{adpn}) whereas experimental analysis of prehistory
probability is only successful for ADP map (\ref{adpn}).

Because there is no an unique escape path for the OGY map, it is
impossible to apply the algorithm described above for controlling
stabilization failures. We note however that, since we know the
dynamics of the fluctuational trajectories, it is still possible
to realize control of the fluctuations by using another approach.
For example, a control force can be added whenever the system
comes to the vicinity of the point $x_1^*$. In this case, however,
the size of the vicinity and the magnitude and form of the control
force are ill defined.

For stabilization of the ADP map, the opposite situation applies:
there exist an unique optimal path and a corresponding optimal
force. Consequently, we can realize a procedure for the control of
large fluctuations. It is similar to that described above for
local control. We monitor trajectories of the map (\ref{adpn}) to
identify the large deviation ($\langle x_n \rangle$, $n=-1$ in
Fig.~7) and in the next iteration we add the control force $u_n= -
\langle \xi_n \rangle$, $n=0$. The dependences on noise intensity
$D$ of the mean time $\langle\tau \rangle$ between stabilization
failures in the absence and in the presence of control are shown
in Fig.~12(a). The dependence of $\langle\tau \rangle$ on the
amplitude of the control force is shown in Fig.~12(b). The
suggested control procedure is evidently effective.

\section*{Conclusion}

We have considered noise-induced failures in the stabilization of
an unstable orbit, and the problem of how to control such
failures. In our investigations, they correspond to large
deviations from stable points. We have examined two types of
stabilization, local and global, and therefore analyzed
fluctuational deviations of different size. We have shown that,
for local stabilization, noise-induced failures can be analyzed
effectively in terms of linearized noisy maps.

Large noise-induced deviations from the fixed point in
one-dimensional maps have been analyzed within the framework of
the theory of large fluctuations. The key point of our
consideration is that the dynamics of the optimal path, and the
optimal fluctuational force, correspond directly to stabilization
failures. We have applied two approaches -- experimental analysis
of the prehistory probability distribution and the solution of the
boundary problem for extended maps -- to determine the optimal
path and the optimal fluctuational force, and we have compared
their results.  For local stabilization, the two approaches give
the same results. For global stabilization, however, the solution
of the boundary problem enabled the optimal path and optimal
fluctuational force to be determined for both the OGY and ADP
maps, whereas investigation of fluctuations' prehistory gave the
optimal path and force for the ADP map only.

A procedure for the control of large fluctuations in
one-dimensional maps has been demonstrated. It is based on the
control concept developed in \cite{Smelyanskiy:97} for continuous
systems. We have introduced an additional control scheme which
significantly improves the stabilization of an unstable orbit in
the presence of noise. It was successful for the ADP method of
stabilization, and problematic for the OGY method. We have shown
that the control procedure has limitations connected with the
existence of unique optimal path and the presence of long time
prehistory of large fluctuation. The relationships between the
large fluctuation dynamics and the control procedures are
summarized in Table I.

The considering of the control problem is relevant to a continuous system 
which has one-dimensional curve in the Poincare section, for example, 
to the Rossler system. For such a continuous system we can formulate 
the task of control as a control in discrete time moments (moments of 
intersection of the Poincare section) by using impulse actions. Intervals 
between the moments were used to calculate and to form a control 
force. Note, that the similar approach is wide used in control 
technique.

The present control approach has limitations which consist of 
necessity of studying the fluctuational dynamics of systems before 
considering of the control problem. Such studying can be done with use of
extended maps of system, if it is known a system model, and/or 
experimentally by fluctuation prehistory analysing. For the local 
stabilization a system model can be easily written down by determination
of eigen-value of a stabilized unstable point; there are a lot of 
effective methods to do it \cite{RMP}. For the global stabilization 
there is no such a way and we need to investigate the fluctuation 
prehistory. Our investigations have shown that in this case we can met 
problems of determination of the control force.  Indeed, we have shown 
that for the global stabilization of OGY map there are several 
most-probable escape paths with, practically, the same energy. As the
result, a real escape path can be a combination of the different 
most-probable paths, so an escape trajectory does not go along a defined 
path as for the ADP map. Moreover, we can not determine the fluctuational 
force and, consequently, the control force, since we use for it averaging 
fluctuational trajectories which follow along an unique path. So, the 
control procedure is inapplicable. It is obviously, that for succesful 
control we have to change the control strategy. For example, we can try 
to predict a fluctuational action locally whereas now we
try to know the full fluctuational dynamics. The local prediction can be 
based on a combination of real time prehistory analysis and reconstruction
of the extended system \cite{Smel_PC}.

Additionally, noise-induced escape through fractal boundaries has
been studied in a one-dimensional map. It was found that
fluctuational motion across fractal basins has a non-activation
character. It was also established that there are several optimal
escape paths from the fixed point of the OGY map (\ref{ogyn})
whereas, for the ADP map (\ref{adpn}), the escape path is unique.
We infer that the existence of several paths in the OGY map
(\ref{ogyn}) is connected with the fact that the stable manifold
of the boundary point $(x_1^*,0)$ goes to the fixed point
$(x^*,0)$.

\begin{acknowledgments}
We thank D.G.\ Luchinsky for useful and stimulating discussions
and help. The research was supported by the EPSRC (UK) and INTAS
01-867.

\end{acknowledgments}

\begin{table}
\begin{tabular}{|p{2cm}|c|c|c|c|}
\hline
&
\multicolumn{4}{c|}{Type of stabilization} \\
\cline{2-5}
&  ADP  &  OGY  &  ADP   &  OGY   \\
&  local & local &  global  &  global  \\
\hline
Unique Optimal Path & \ \ \ \ X \ \ & \ \ \ \ X \ \ & \ \ \ \ X \ \ & \\
\hline
Long Prehistory & \ \ \ \ X \ \ & & \ \ \ \ X \ \ & \ \ \ \ X \ \ \\
\hline
Successful Control &\ \ \ \ X \ \ & & \ \ \ \ X \ \ &\\
\hline
\end{tabular}
\caption{The relationships between the dynamics of fluctuational
paths and the control procedures.}
\end{table}

\begin{figure}[f]
\label{fig1}
\includegraphics[height=2.5in]{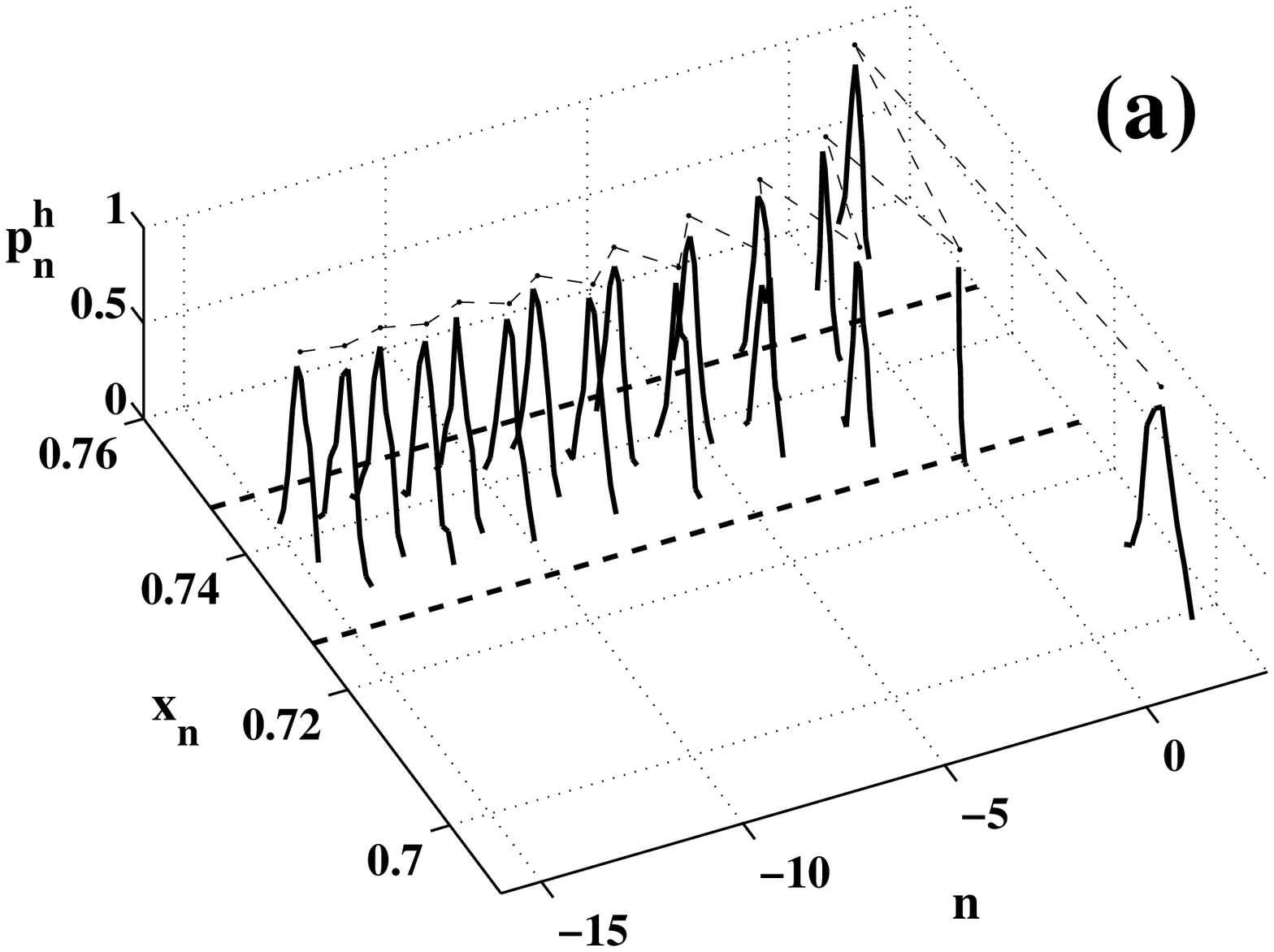}~~~~
\includegraphics[height=2.5in]{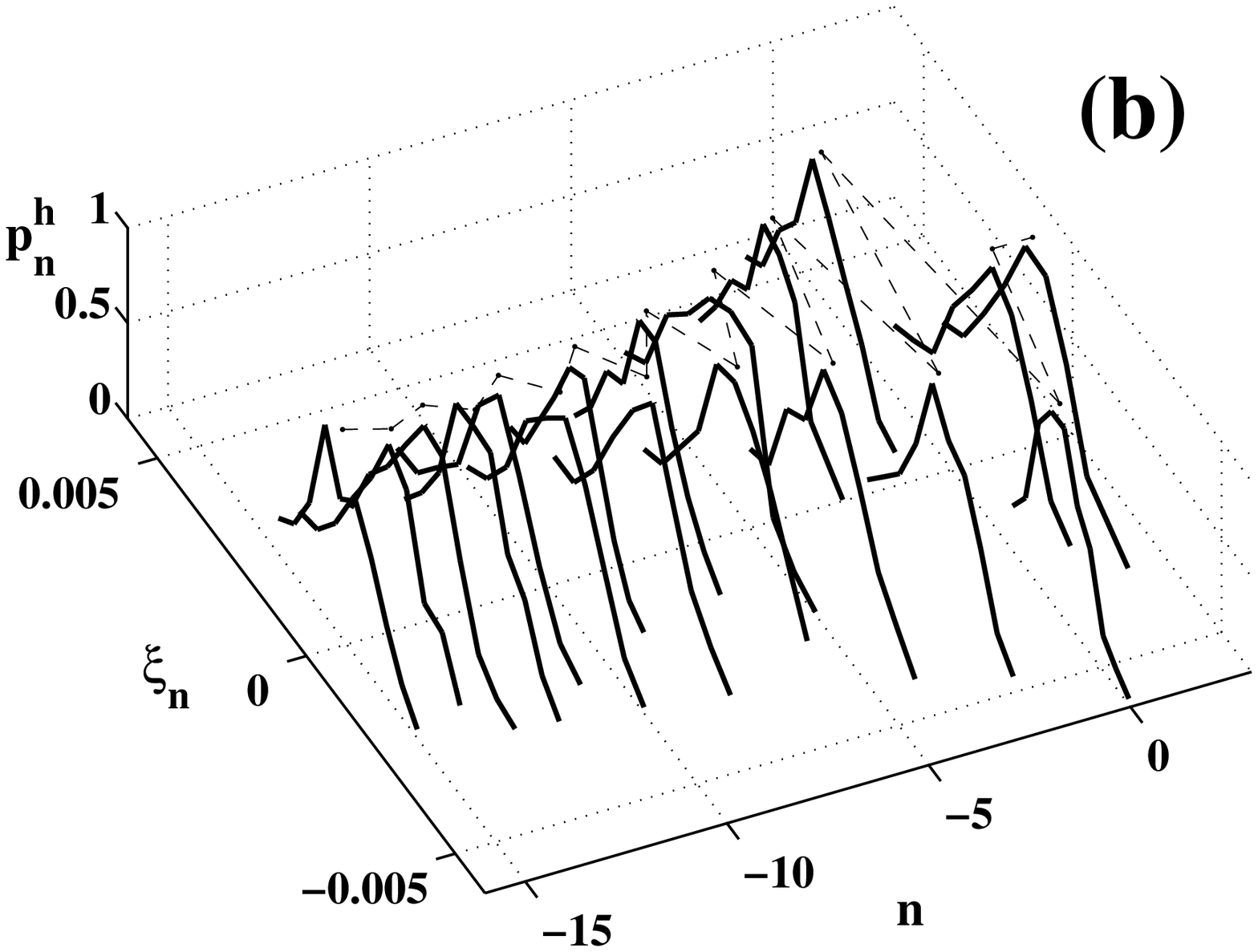}~~~~
\caption{PPDs $p_h^n$ of the exit trajectories (a) and noise
realizations (b) of ADP map for the boundary $(x^*- \epsilon)$.
The thick dashed lines indicate   $\epsilon$-region of
stabilization. The thin dashed lines connect maxima of PPDs. The
noise intensity is $D=0.0011$. }
\end{figure}

\begin{figure}[f]
\label{fig2}
\includegraphics[height=2.5in]{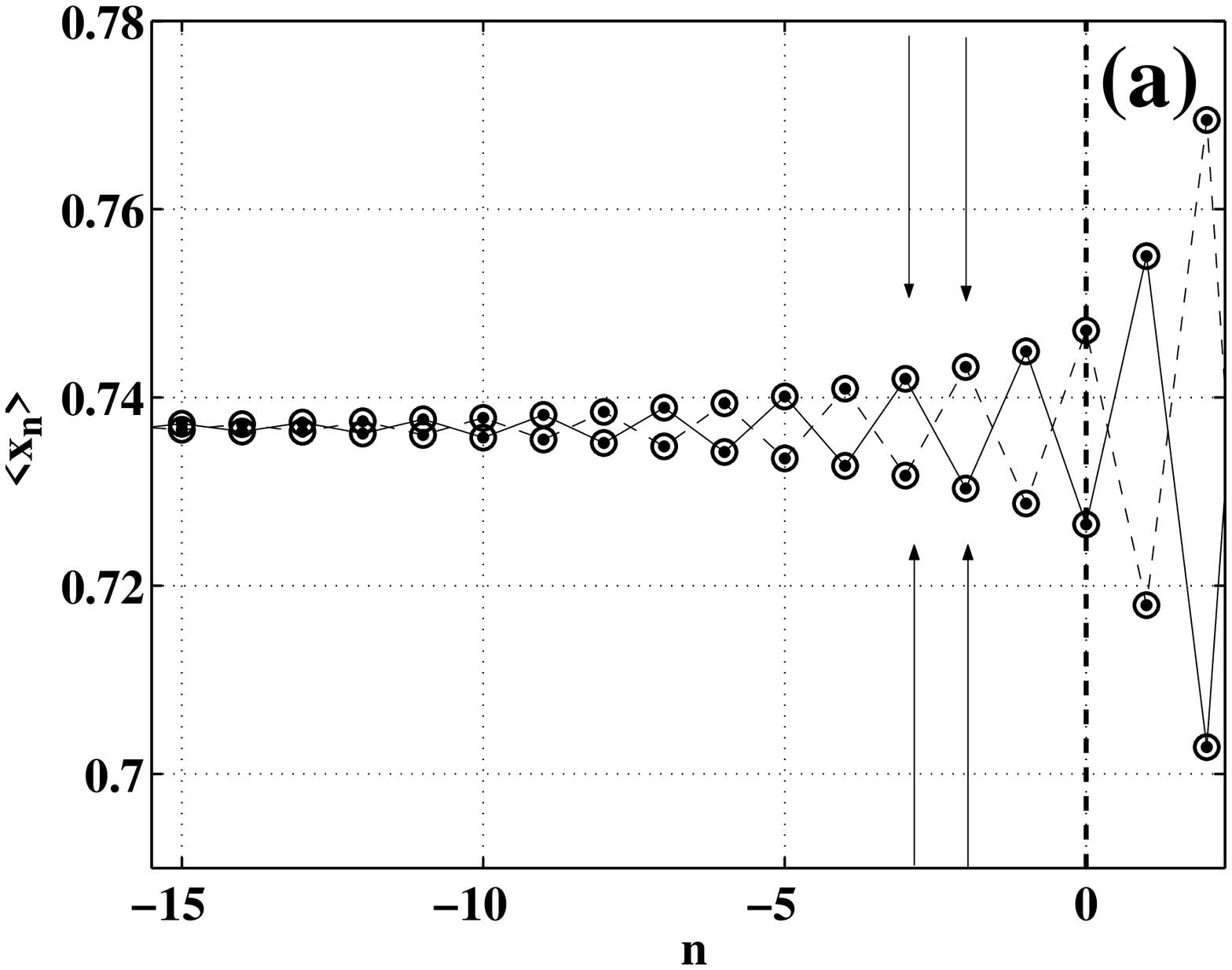}~~~~
\includegraphics[height=2.5in]{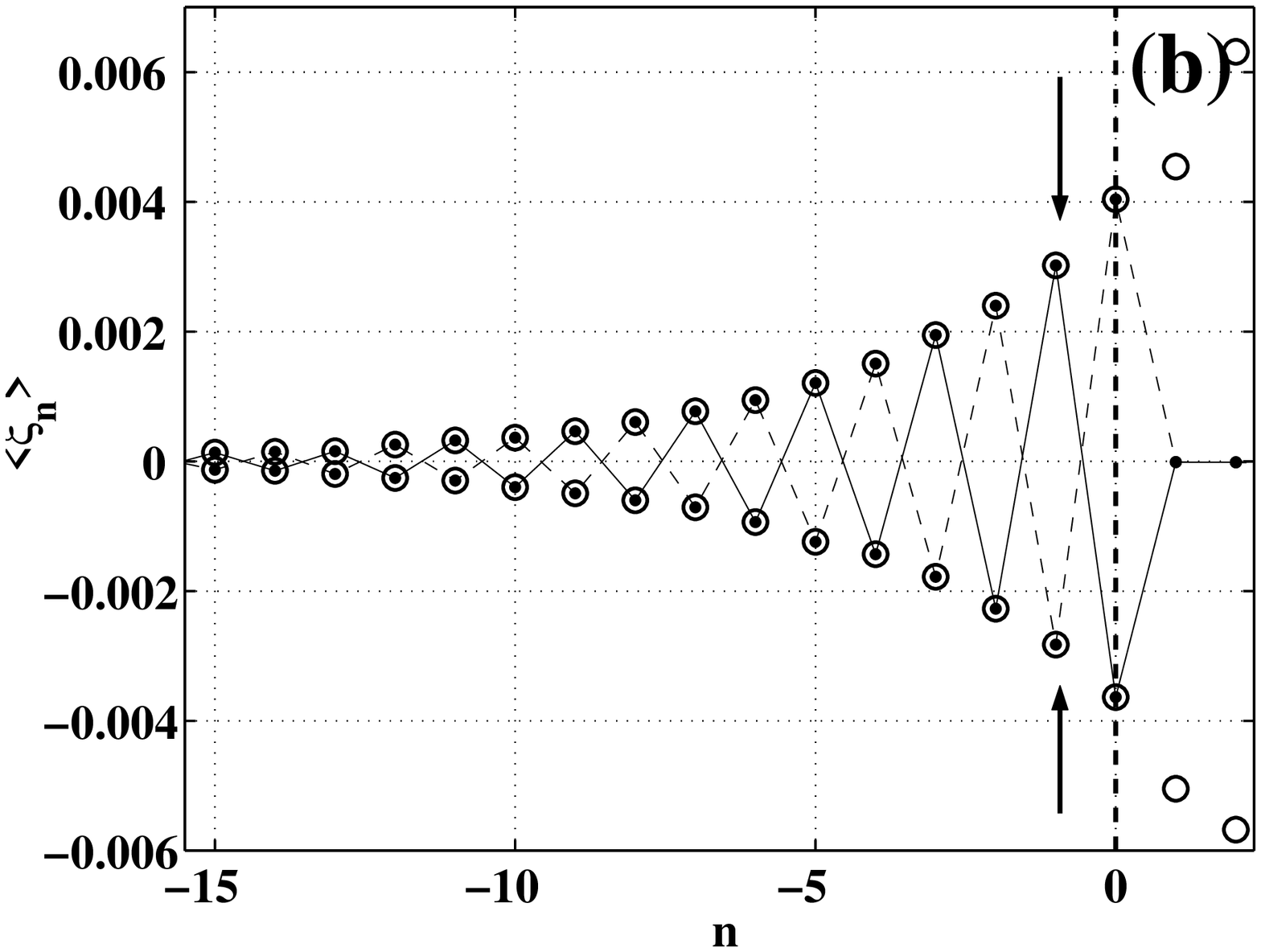}
\caption{ The optimal paths (a) and the optimal forces (b) for
exit through the boundary $(x^*-\epsilon)$ (solid line) and the
boundary $(x^*+\epsilon)$ (dashed line) for ADP map. Circles
indicate the optimal paths and forces obtained by solving the
boundary problem for the linearized extended map (\ref{aux}). The
optimal paths and forces used in the control procedure are marked
by arrows.}
\end{figure}

\begin{figure}[f]
\label{fig3}
\includegraphics[height=2.5in]{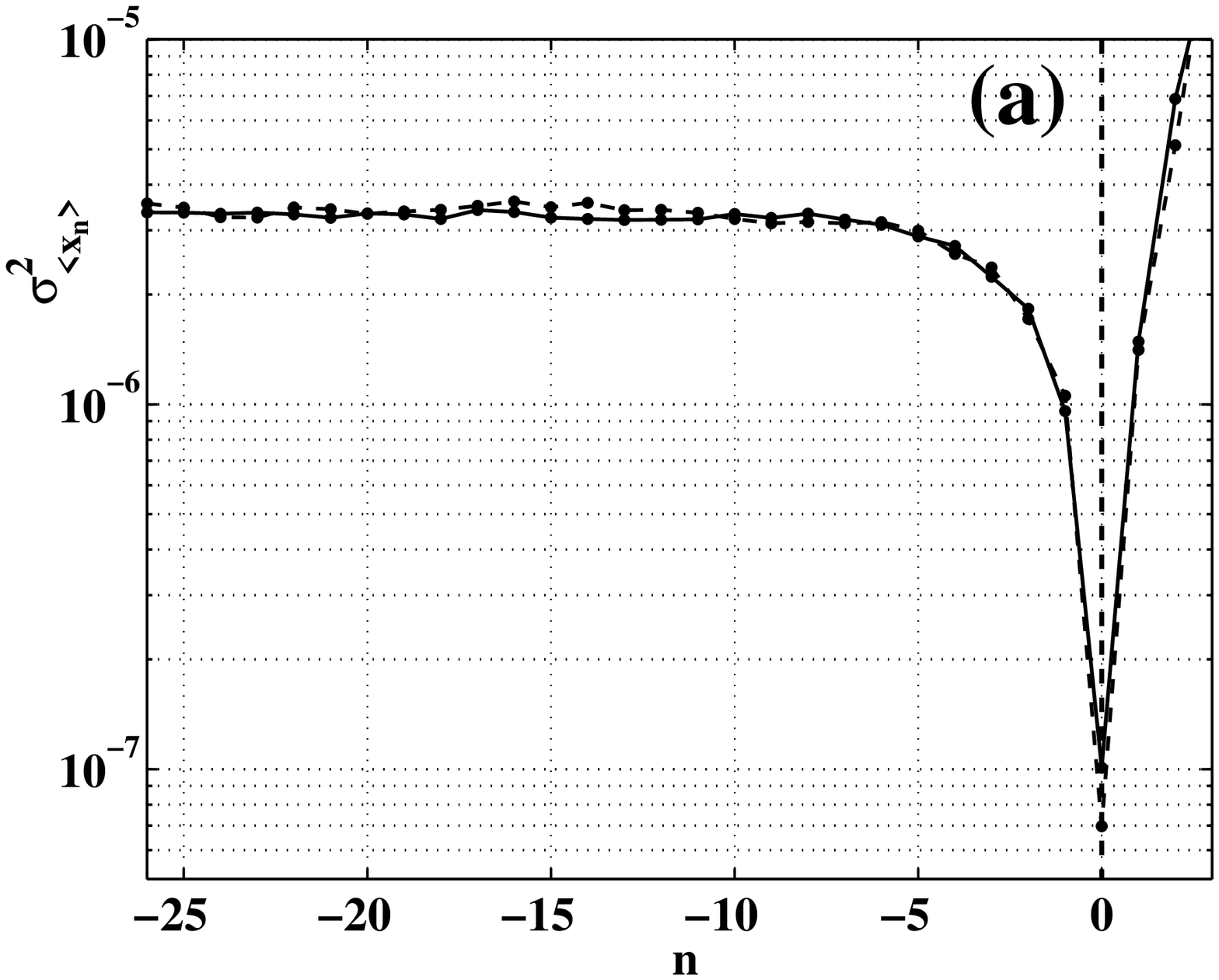}~~~~
\includegraphics[height=2.5in]{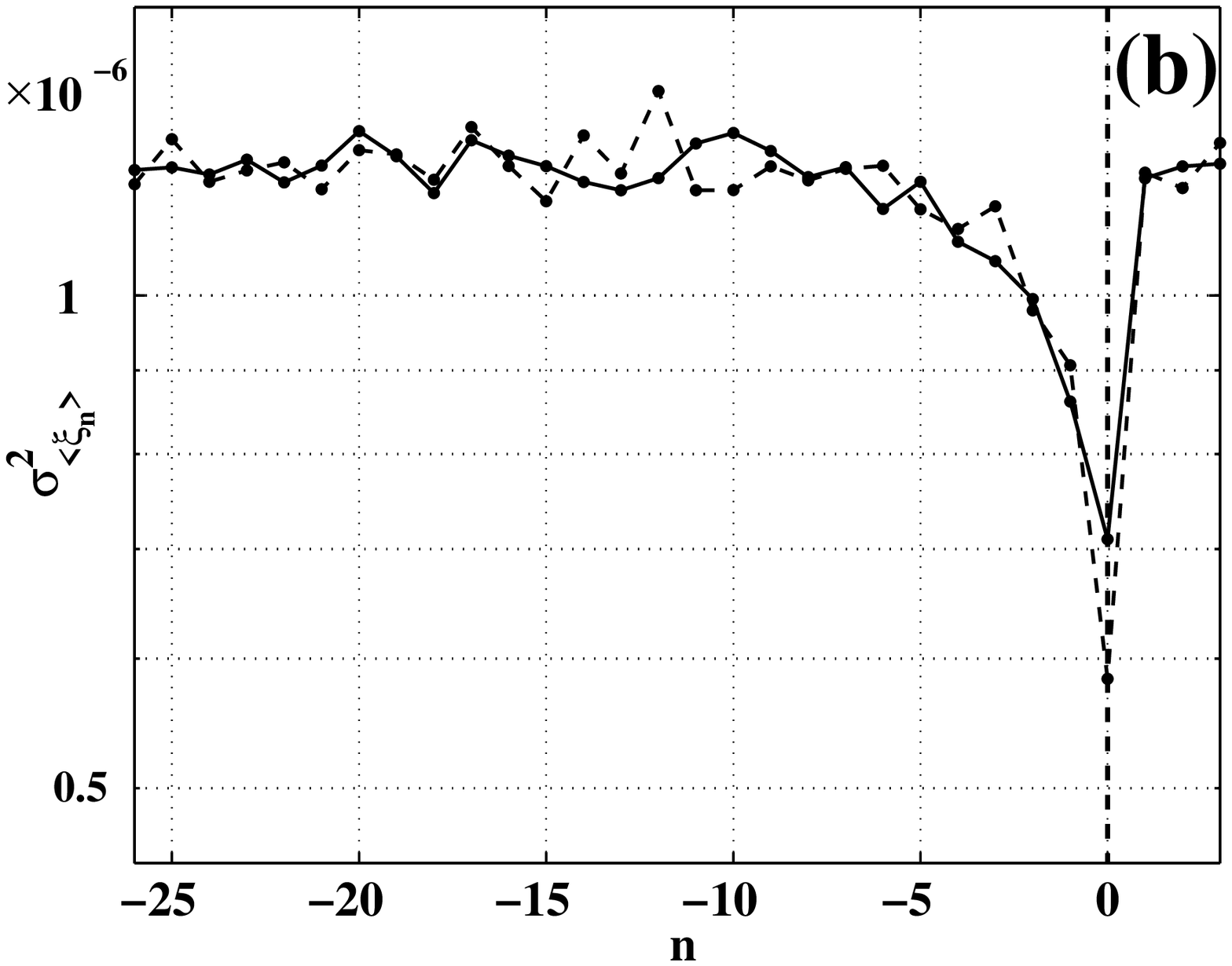}
\caption{(a) The dispersion of the exit trajectories, and (b) the
dispersion of the corresponding noise realizations for exit
through the boundary $(x^*-\epsilon)$ (solid line) and the
boundary $(x^*+\epsilon)$ (dashed line) for the ADP map. }
\end{figure}

\begin{figure}[f]
\label{fig4}
\includegraphics[height=2.5in]{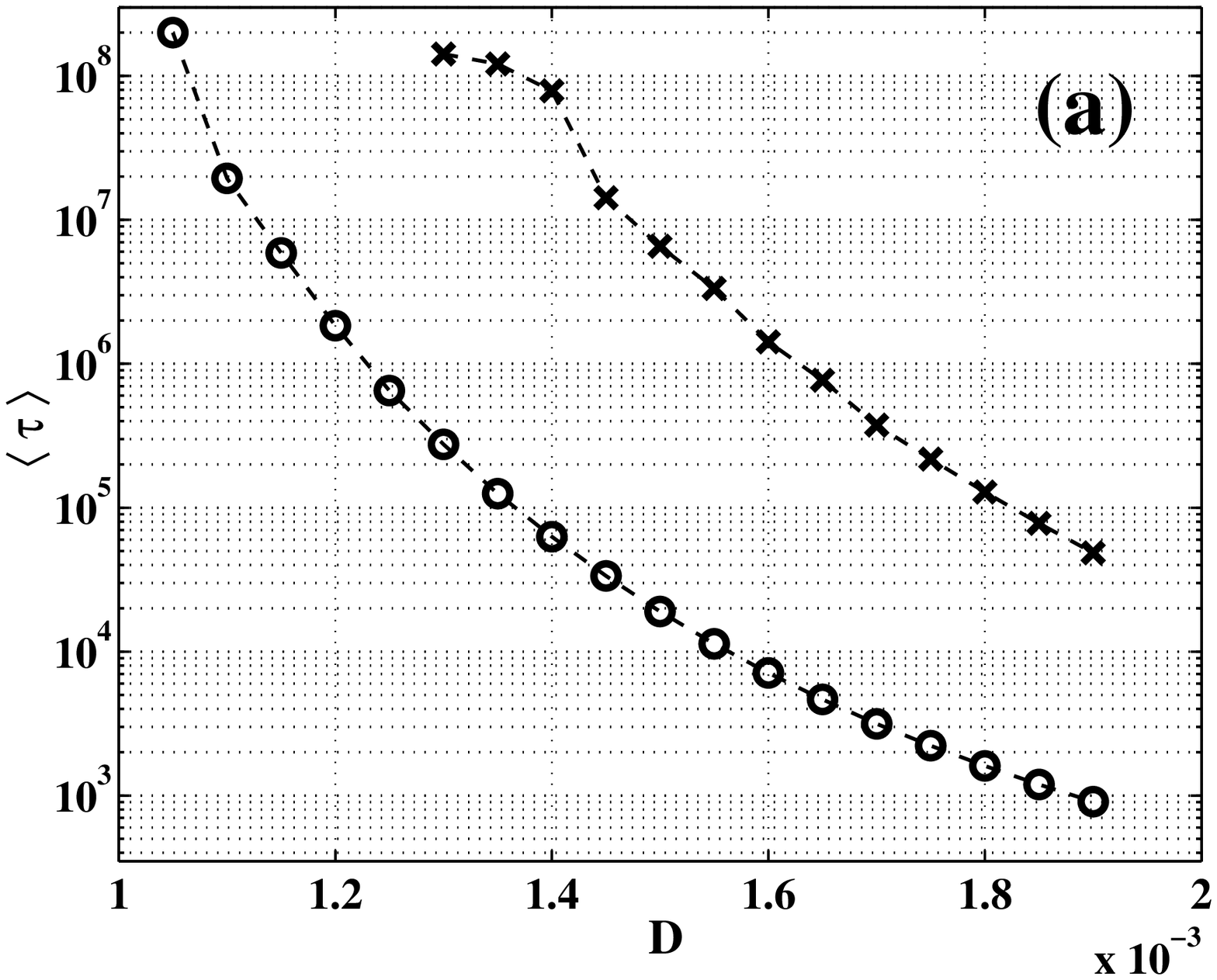}~~~~
\includegraphics[height=2.5in]{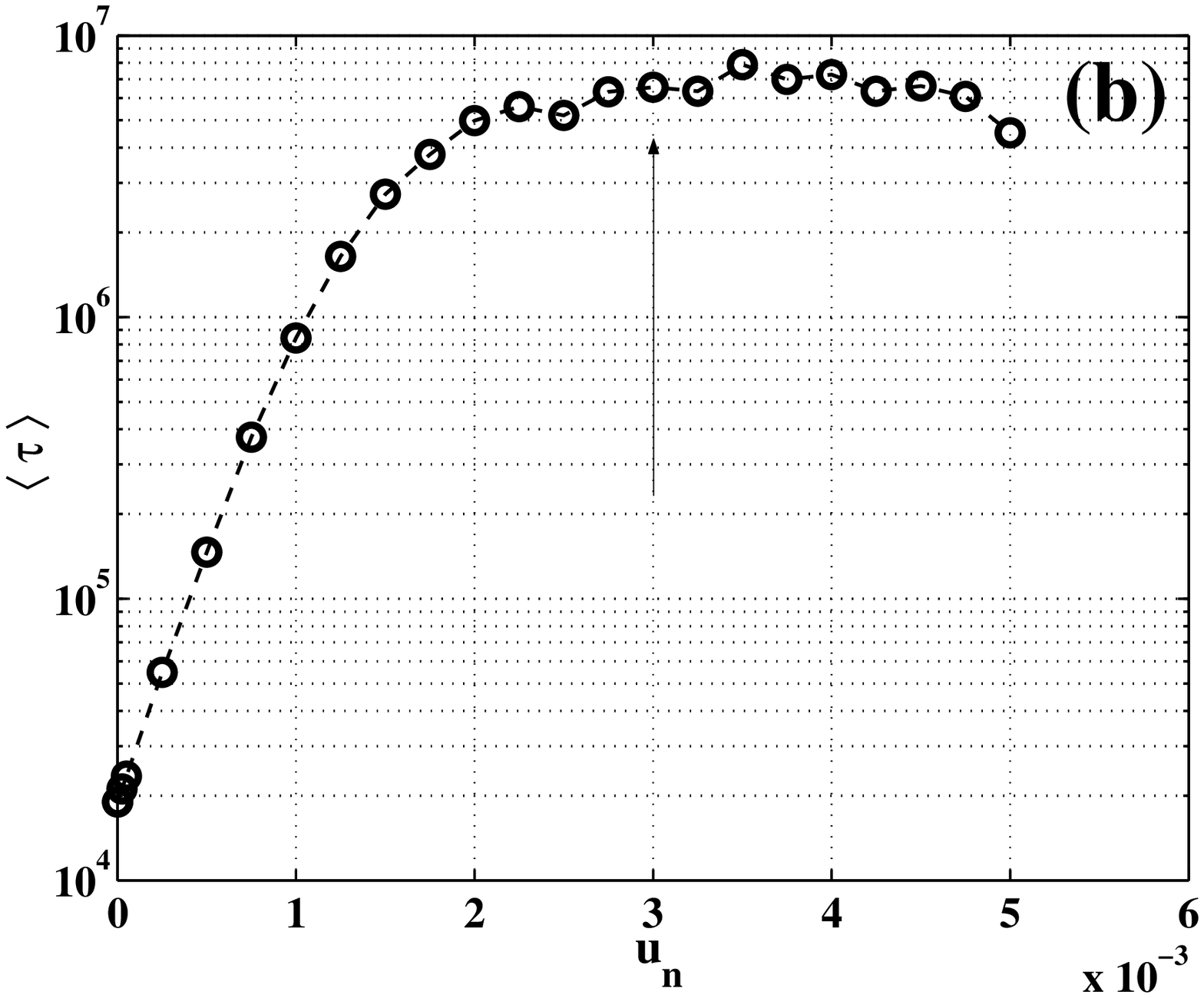}
\caption{(a) The dependences of mean time $\langle \tau \rangle$
between stabilization failures on noise intensity $D$ in the
absence (circles)  and in the presence (crosses) of the control.
The size of the stabilization region is $\epsilon=0.01$. (b) The
dependence of the mean time $\langle \tau \rangle$ on the
amplitude of the control force $u_n$ is presented for the ADP
method. The value of $\langle \tau \rangle$ corresponding to the
optimal fluctuational force is marked by the arrow.
 }
\end{figure}

\begin{figure}[f]
\label{fig5}
\includegraphics[height=2.5in]{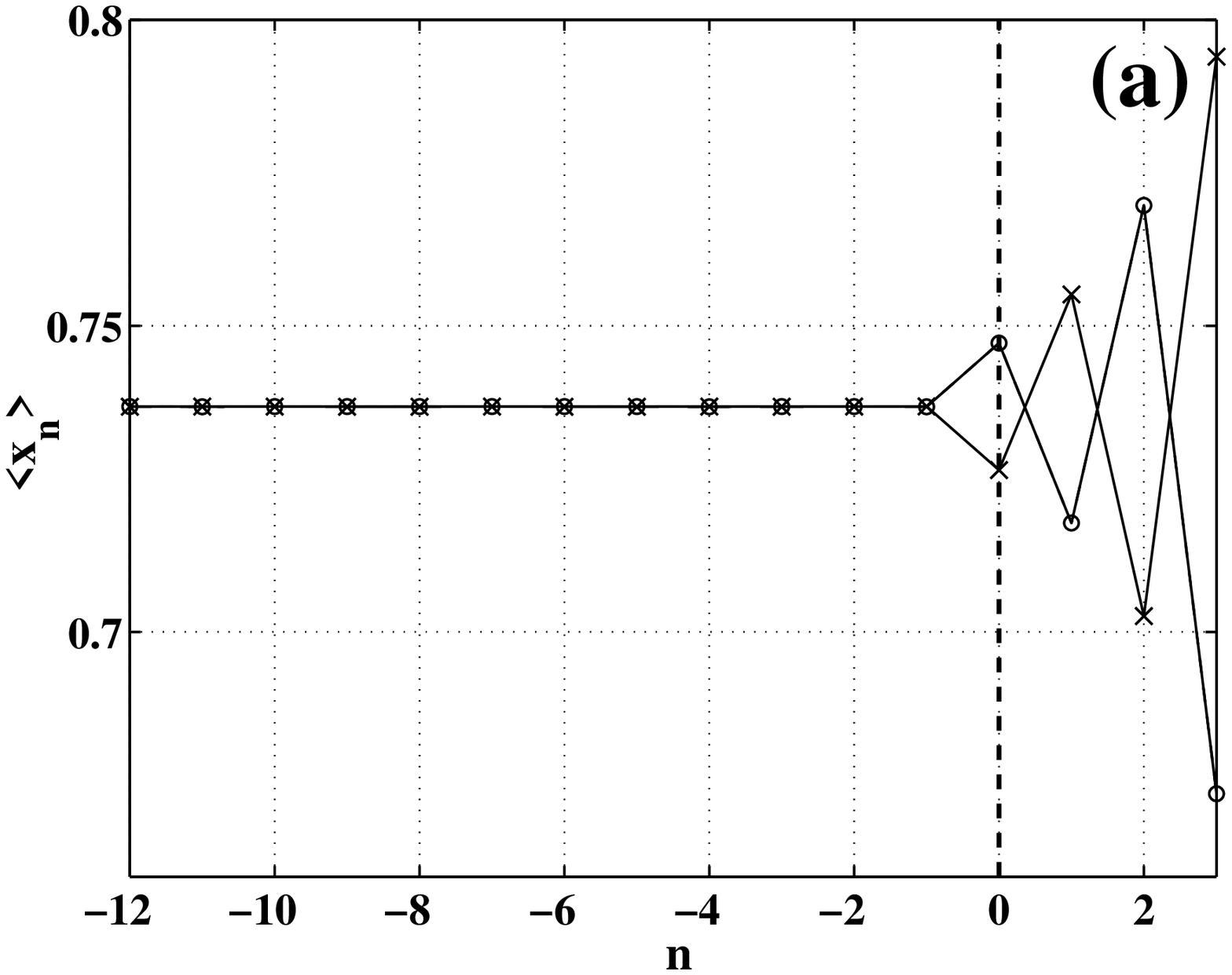}~~~~
\includegraphics[height=2.5in]{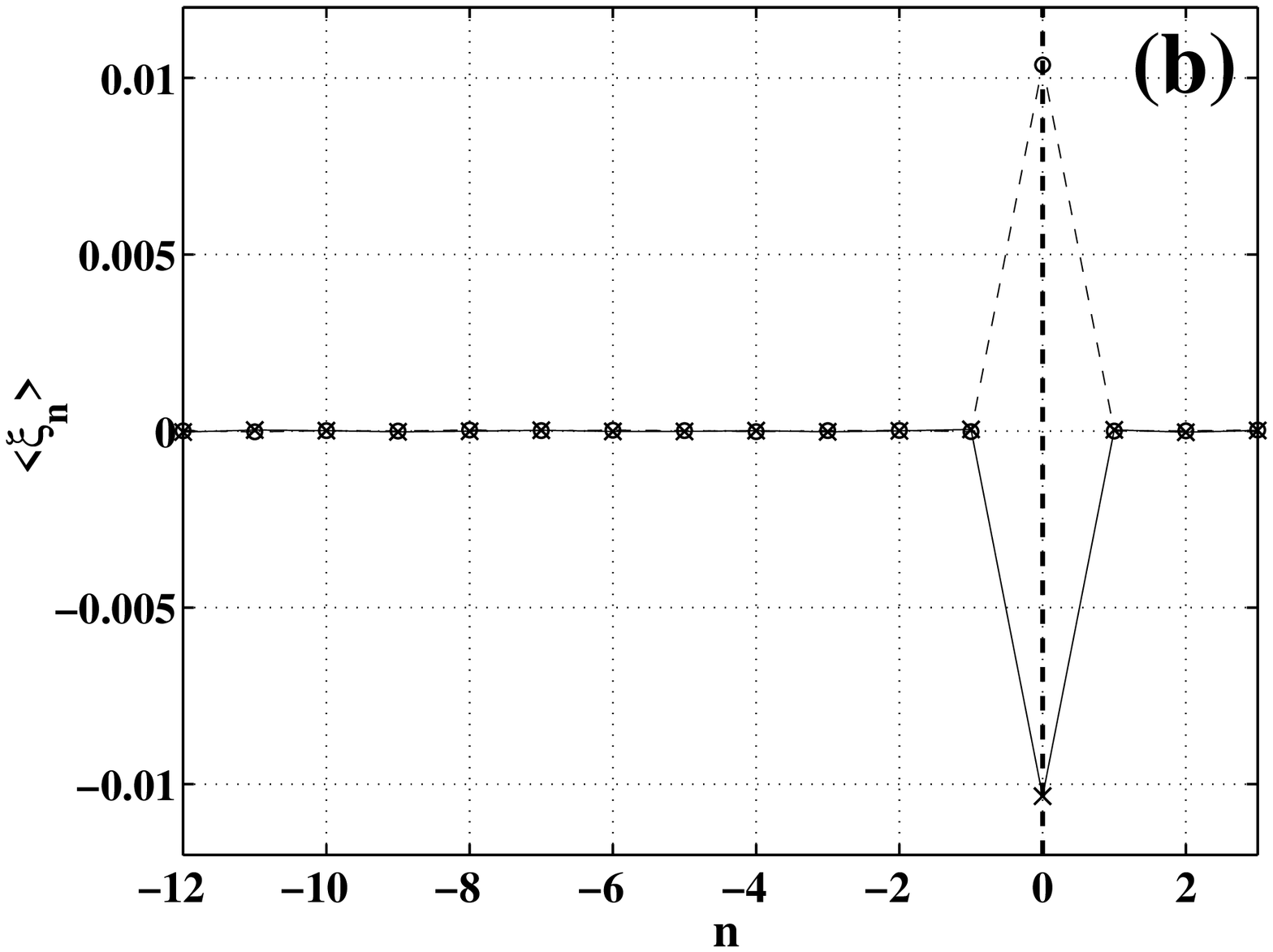}~~~~
\caption{ For the OGY map, the optimal path (a) and the optimal
force (b) are shown for exit through the boundary $(x^*-\epsilon)$
(crosses) and the boundary $(x^*+\epsilon)$ (circles). }
\end{figure}

\begin{figure}[h]
\label{fig6}
\includegraphics[height=2.5in]{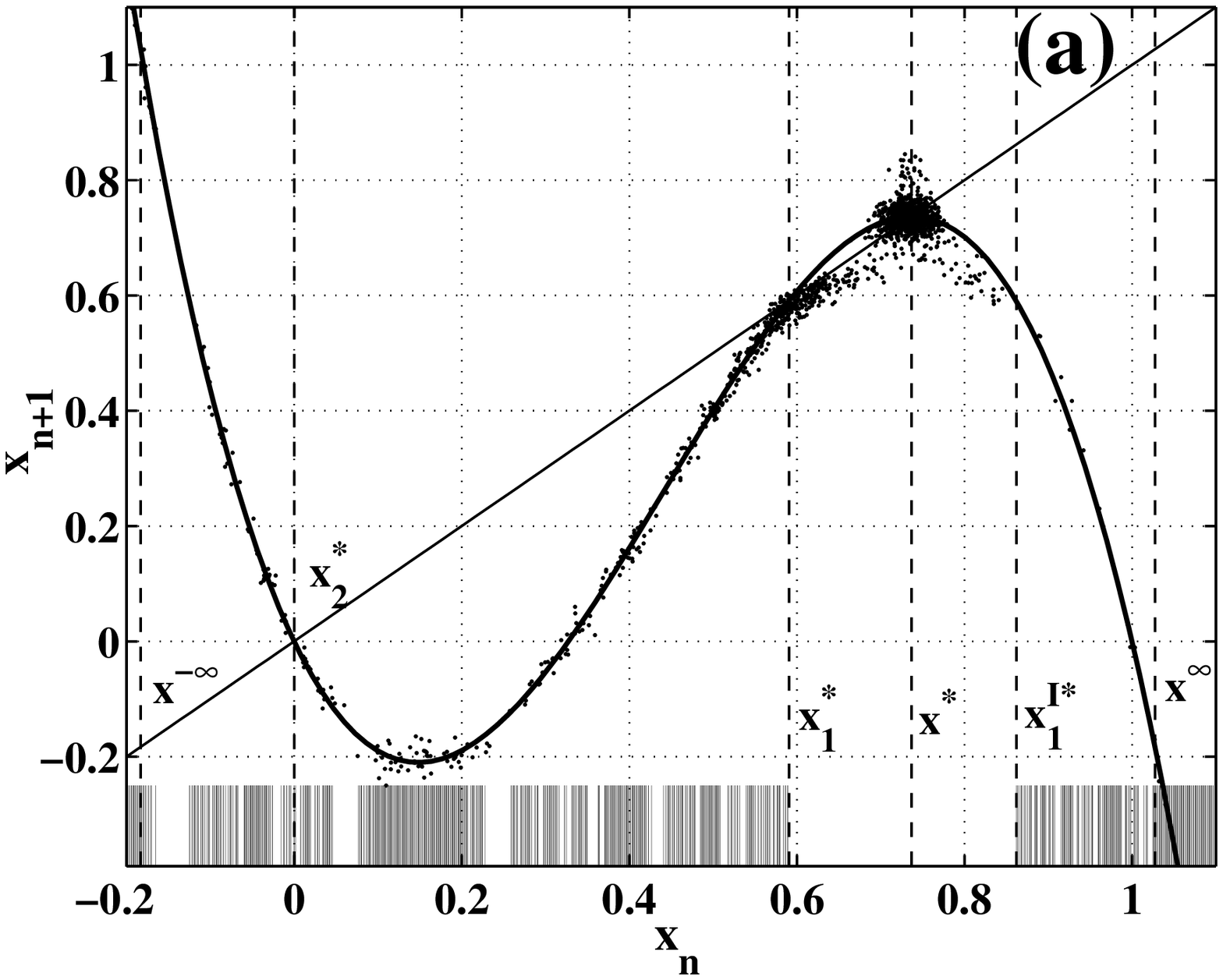}~~~~
\includegraphics[height=2.5in]{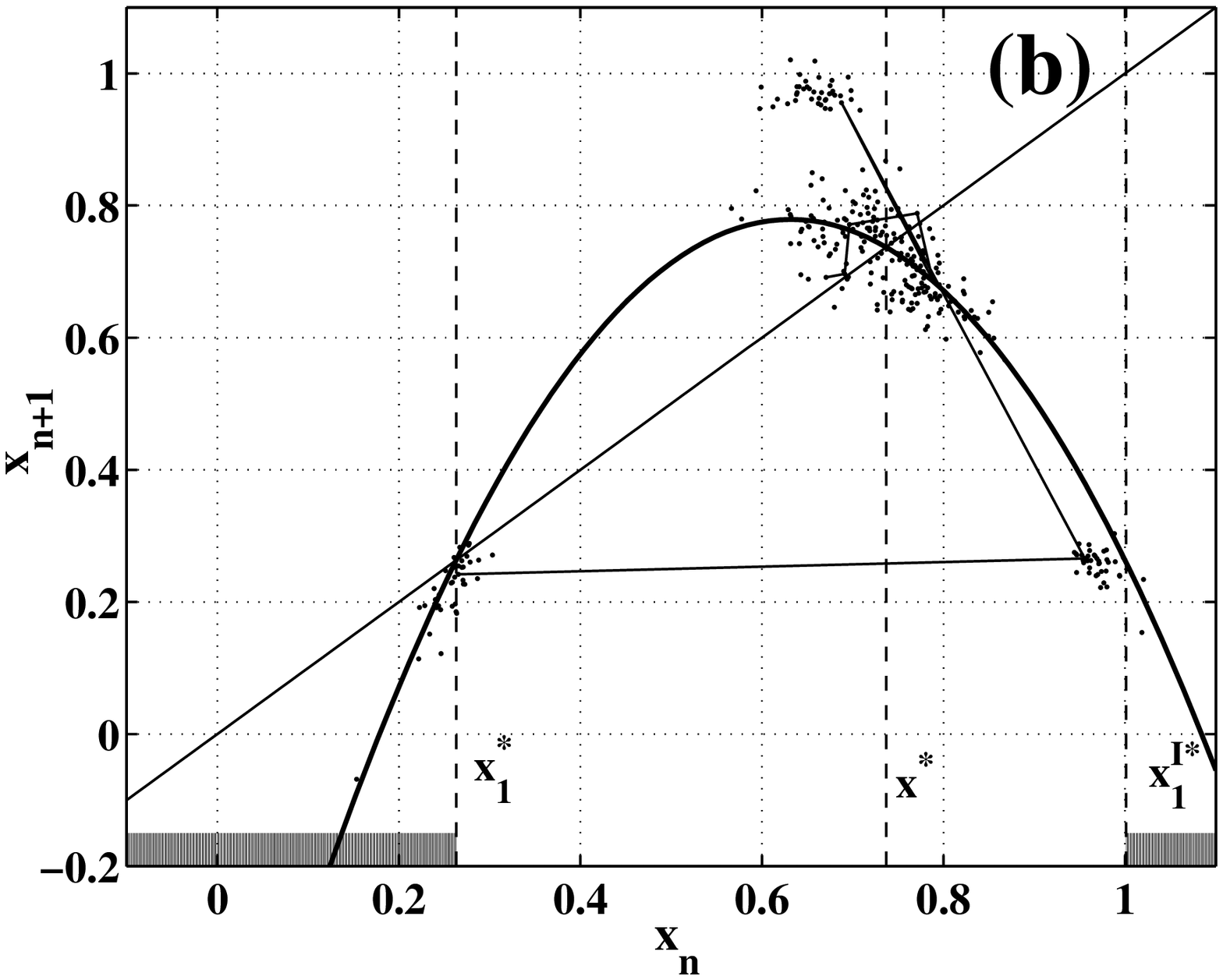}
\caption{ The OGY map (a) and the ADP map (b) on the plane
$(x_n-x_{n+1})$ are shown by the thick solid line. Basins of
attraction of the fixed point $x^*$ (white regions) and the
attractor at infinity (black regions) are shown at the bottoms of
the figures. The dashed lines indicate locations of the fixed
points of the maps and the points defining the basin boundaries.
Escape trajectories are shown by dots. The thin line in figure (b)
corresponds to the optimal path.}
\end{figure}

\begin{figure}[h]
\label{fig7}
\includegraphics[height=2.5in]{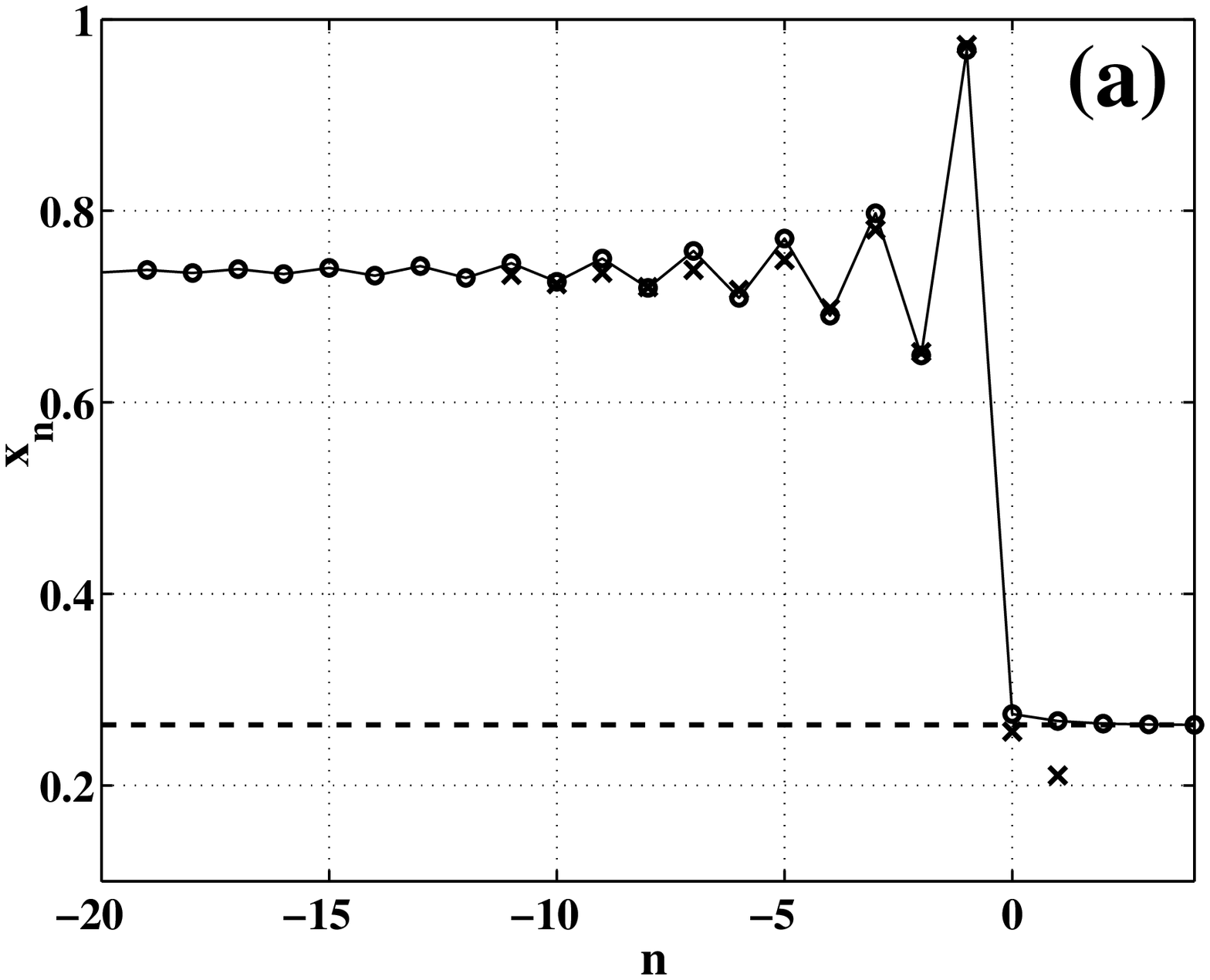}~~~
\includegraphics[height=2.5in]{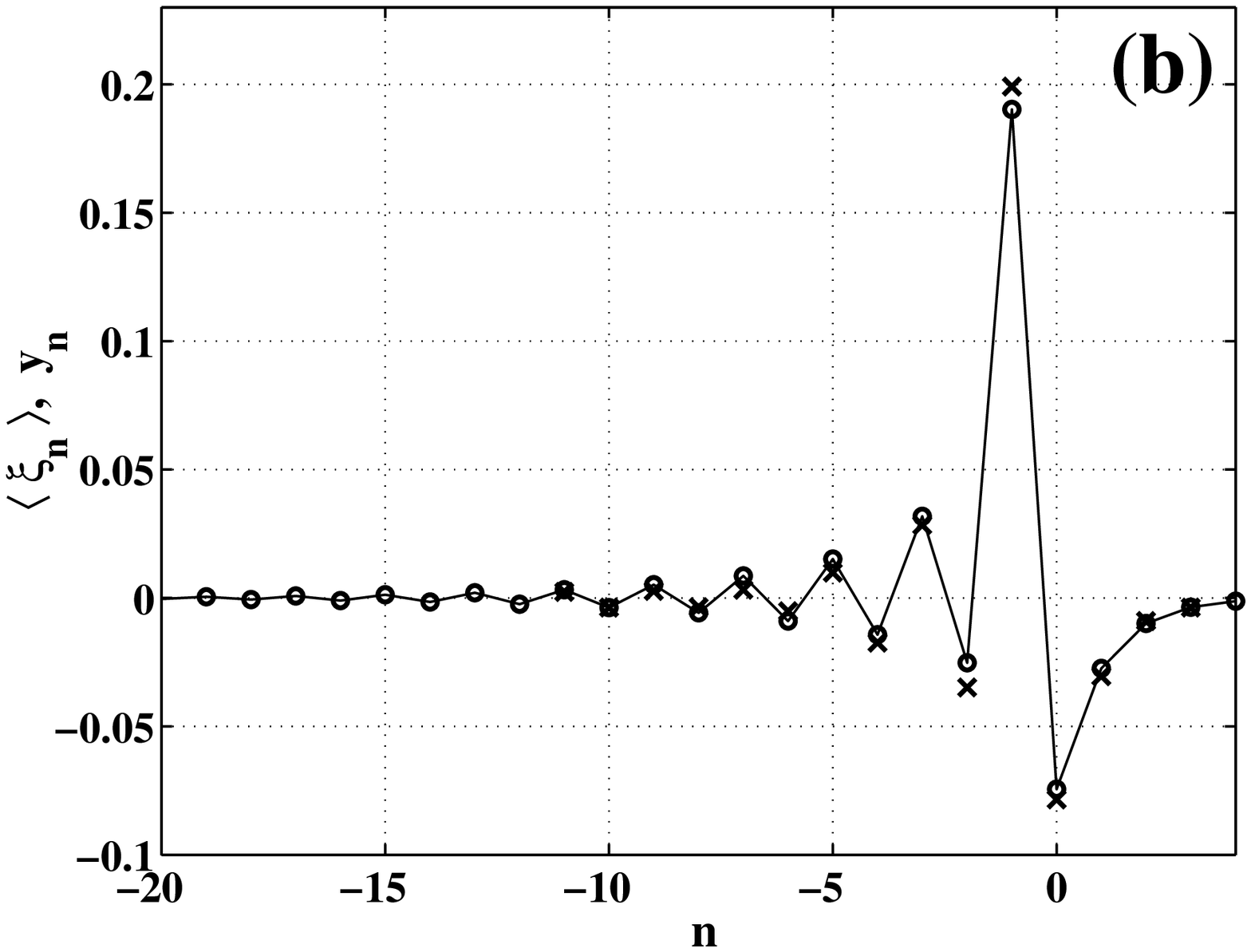}
\caption{ The optimal path (a) and the optimal force (b) obtained
by experimental analysis of the PPD for the ADP map (\ref{adpn})
(crosses) and by solving the boundary problem for the map
(\ref{hamogy}) (circles). }
\end{figure}

\begin{figure}[h]
\label{fig8}
\includegraphics[height=2.5in]{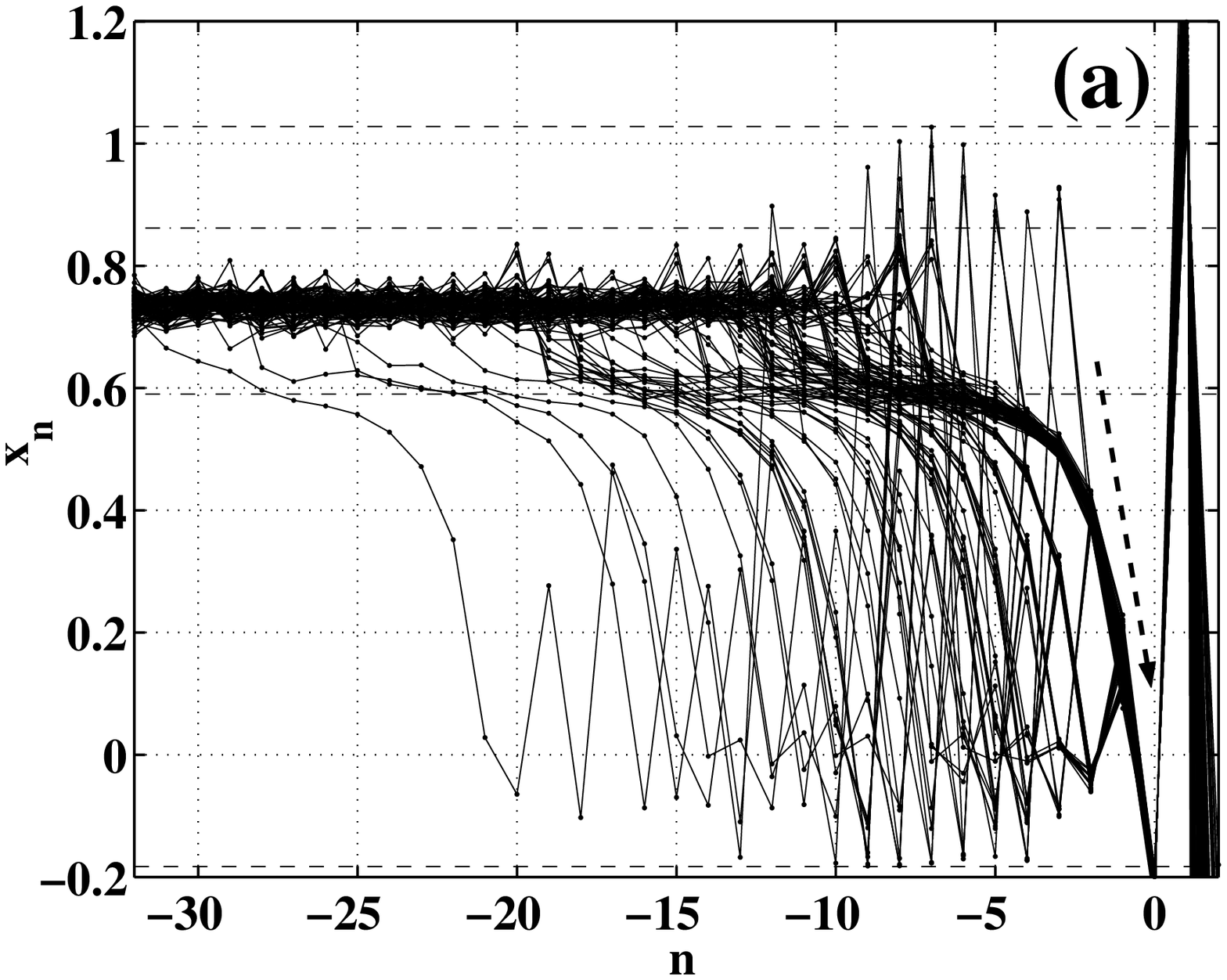}~~~
\includegraphics[height=2.5in]{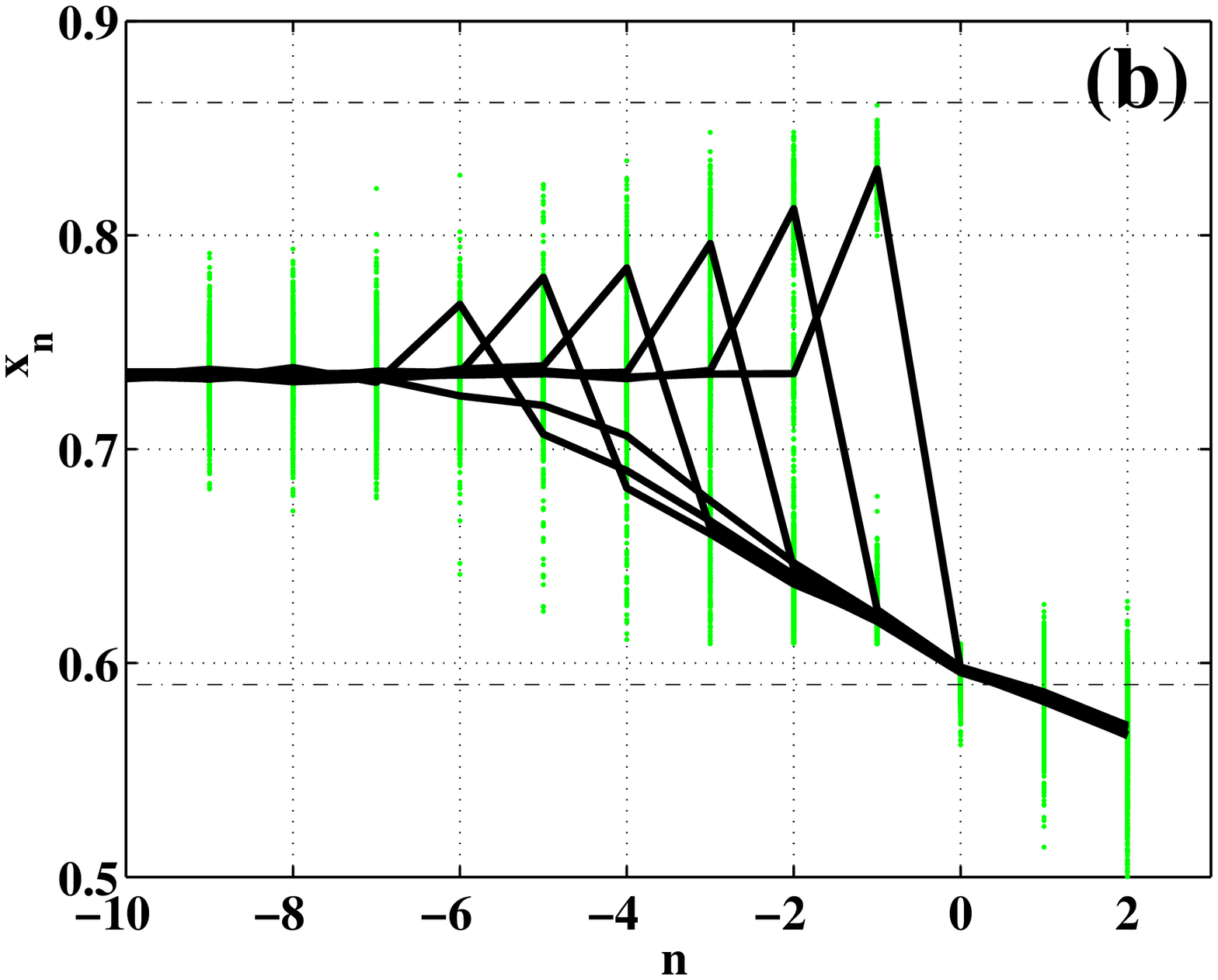}
\caption{ 100 escape trajectories of the OGY map collected in a
vicinity of (a) the point $x^{-\infty}$ and (b) the point $x^*_1$.
The size of the vicinity is defined as the mean square of noise
intensity $D$. The dash-dot lines indicate the location of the
boundary point $x^*_1$ and its pre-image $x^{I*}_1$; the dashed
lines in (a) represent boundaries of the basins with fractal
structure, i.e.\ of the points $x^{-\infty}$ and $x^{\infty}$. The
thick lines in (b) correspond to different escape paths; the grey
dots are coordinates of escape trajectories. The noise intensity
is $D=0.018$. }
\end{figure}

\begin{figure}[h]
\label{fig9}
\includegraphics[height=2.5in]{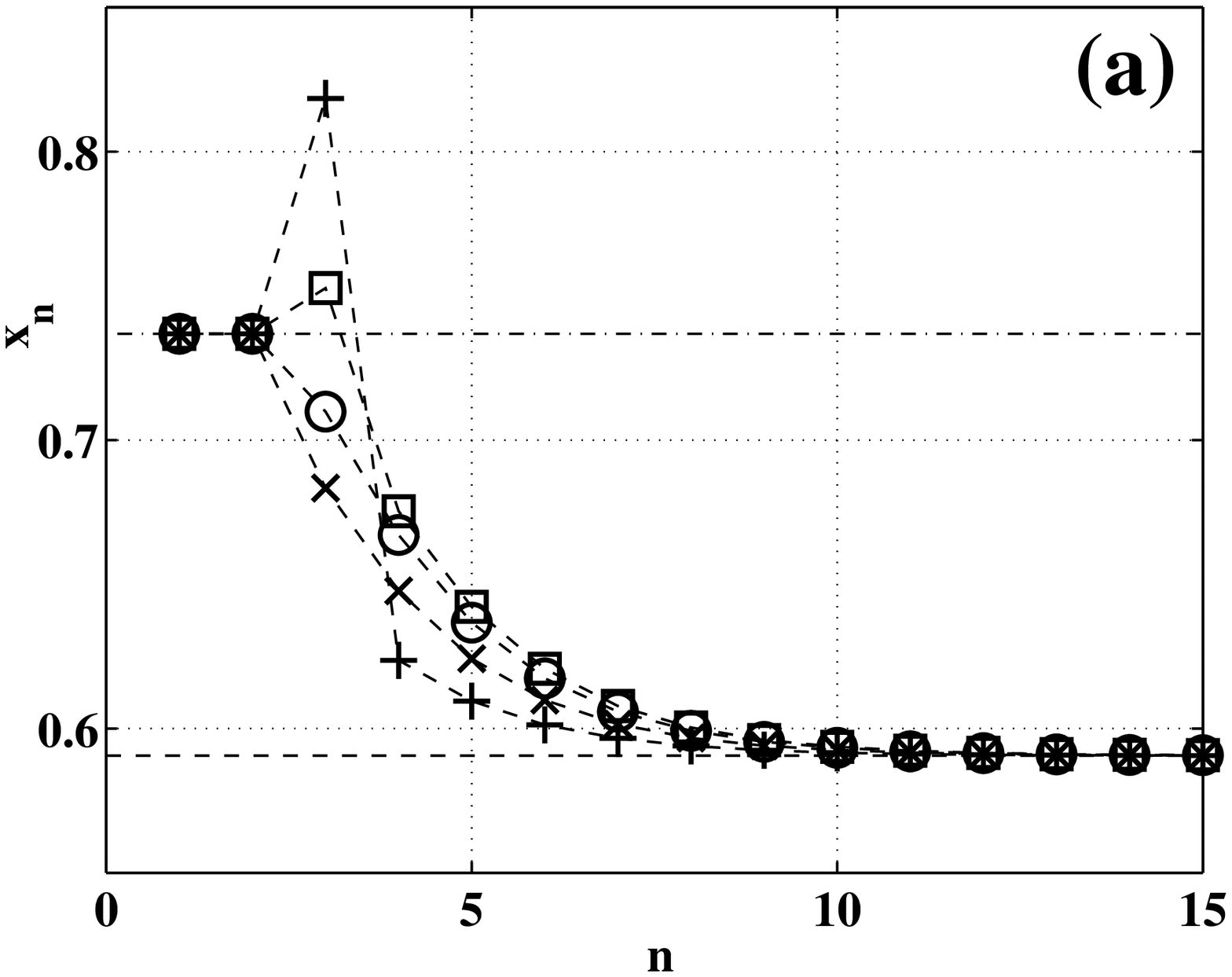}~~~~
\includegraphics[height=2.5in]{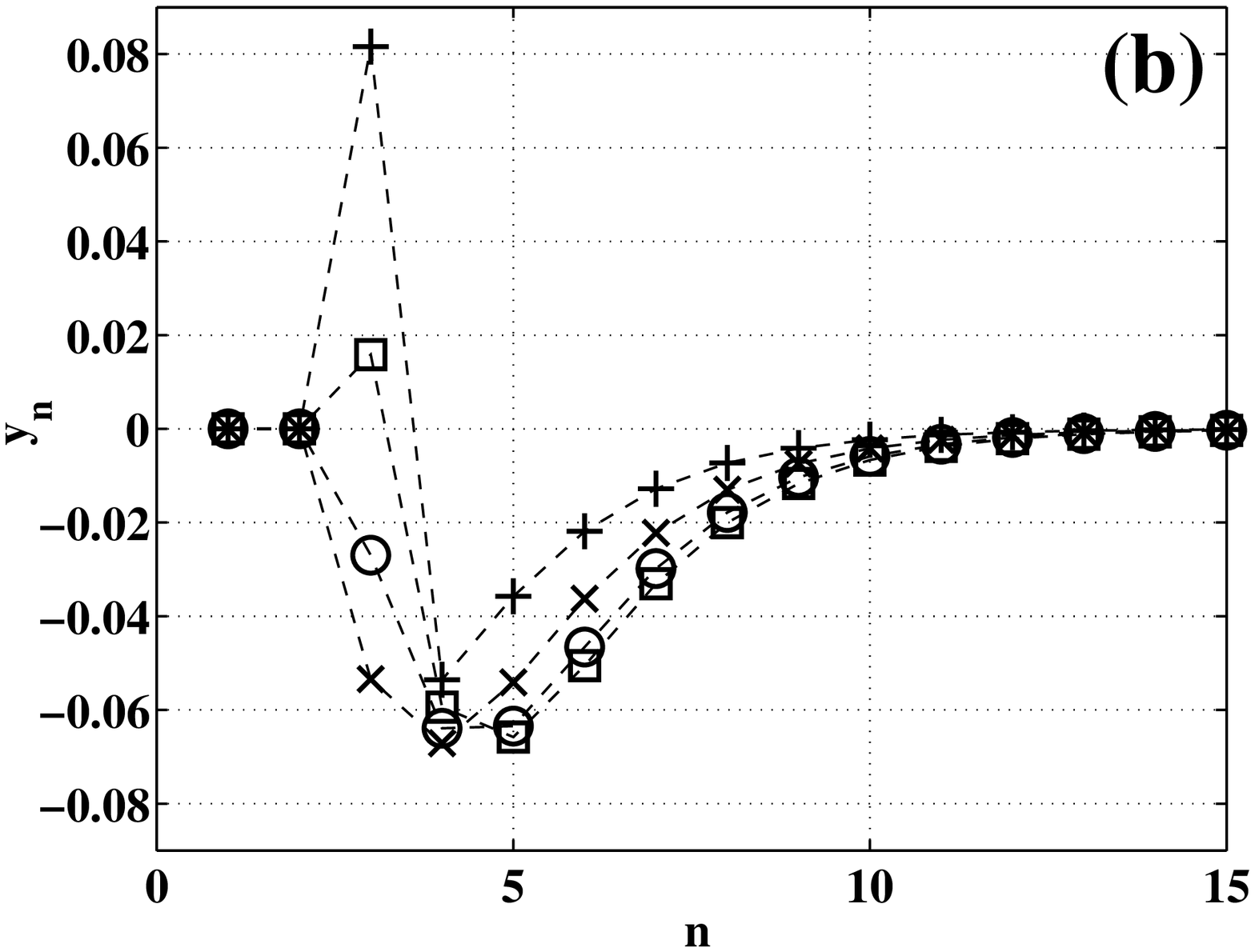}
\caption{The optimal paths (a)  and optimal forces (b) obtained by
solution of the boundary problem for the OGY map. The path $t_1$
is marked with $\circ$, $t_2$ --- $\Box$, $t_3$
---  $\times$, $t_4$ --- $+$. }
\end{figure}

\begin{figure}[h]
\label{fig10}
\includegraphics[height=3in]{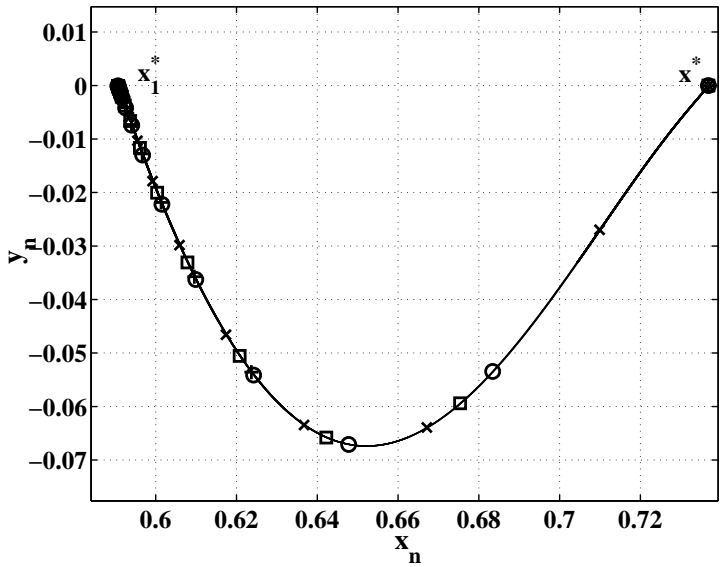}
\caption{ The stable manifold of the point $x^*_1$ of the OGY map.
The symbols indicate the different optimal paths, using the same
coding as in Fig.~9. }
\end{figure}

\begin{figure}[f]
\label{fig11}
\includegraphics[height=2.5in]{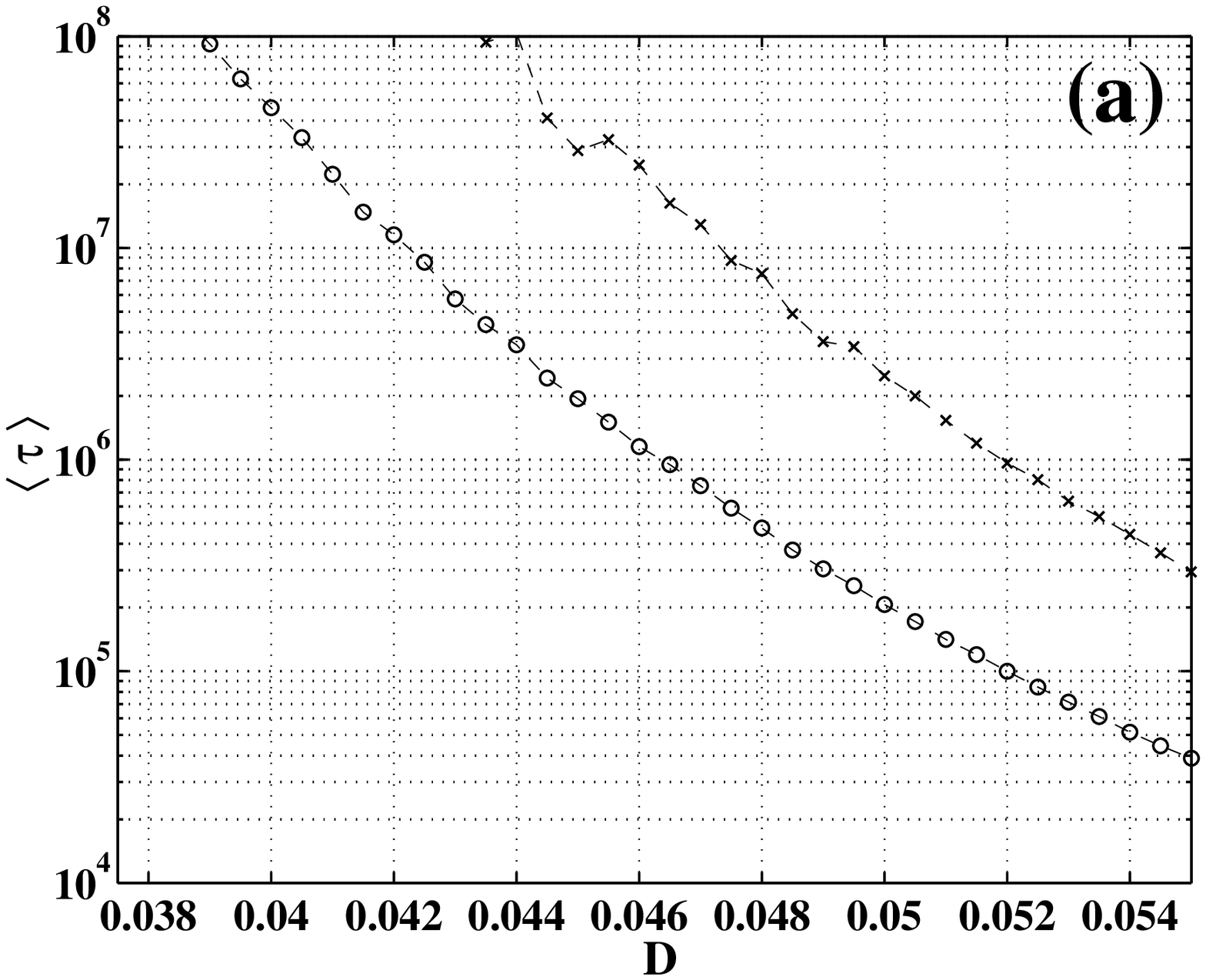}~~~~
\includegraphics[height=2.5in]{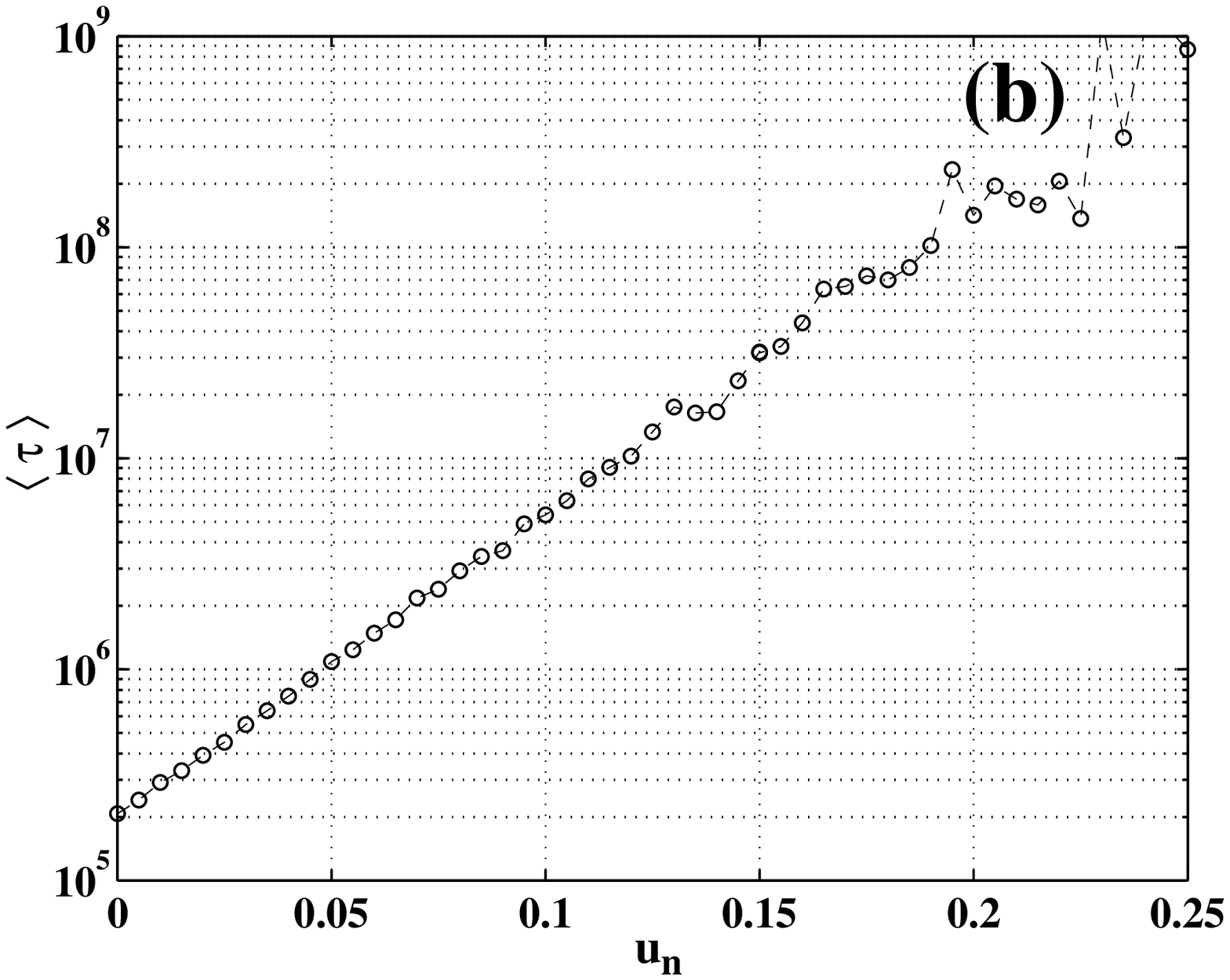}
\caption{(a) Dependences of the mean time $\langle \tau \rangle$
between stabilization failures on noise intensity $D$ in the
absence (circles) and in the presence (crosses) of the control.
The stabilization is global, using the ADP method. (b) Dependence
of the mean time $\langle \tau \rangle$ on the amplitude of the
control force $u_n$.
 } \end{figure}


\begin{thebibliography}{99}
\bibitem{Boccalettia:00}
S.\ Boccalettia et al, Phys.\ Rep.
{\bf 329}, 103 (2000).
\bibitem{Ott:90} E.\ Ott, C.\ Grebogy, J.\ Yorke, Phys.\ Rev.\ Lett. {\bf 64},
1196 (1990).
\bibitem{Bishop:96}
S.~R.\ Bishop and D.\ Xu,
Phys.\ Rev.\ E {\bf 54}, 3204 (1996).
\bibitem{Botina:97}
J.\ Botina and H.\ Rabitz
Phys.\ Rev.\ E {\bf 56}, 3854 (1997).
\bibitem{Collins:95}
D.~J.\ Christini and J.~J.\ Collins,
Phys.\ Rev.\ E {\bf 52}, 5806 (1995).
\bibitem{Graham:91}
R.\ Graham and T.\ Tel, Phys.\ Rev.\ Lett. {\bf 66}, 3089  (1991).
\bibitem{Grassberger:89}
P.\ Grassberger,  J.\ Phys.\ A. {\bf 22}, 3283 (1989).
\bibitem{Dykman:92}  M.~I.\ Dykman et al,  Phys.\ Rev.\ Lett.
{\bf 68}, 2718 (1992).
\bibitem{rt} G.\ Marsaglia, and W.~W.\  Tsang,
 SIAM\ J.\  Sci.\  Stat.\  Comput. {\bf 5}, 349 (1984).
\bibitem{Pontryagin} L.~S.\ Pontryagin, {\it The Mathematical Theory
of Optimal Processes} (Macmillan, 1964).
\bibitem{Whittle:96} P.\ Whittle,
{\it Optimal Control: Basics and Beyond} (Wiley, 1996).
\bibitem{Smelyanskiy:97} V.~N.\ Smelyanskiy and M.~I.\ Dykman, Phys.\ Rev.\ E
{\bf 55}, 2516 (1997).
\bibitem{Rabitz:97} B.~E.\ Vugmeister and H.\ Rabitz, Phys.\ Rev.\ E {\bf
55},
2522 (1997).
\bibitem{Luchinsky:97} D.~G.\ Luchinsky,
 J.\ Phys.\ A {\bf 30}, L577 (1997).
\bibitem{Khovanov:00} I.~A.\ Khovanov et al,   Phys.\ Rev.\ Lett.
{\bf 85}, 2100 (2000).
\bibitem{numeric}
W.~H.\ Press et al. {\it Numerical recipes: the art of scientific computing}
(Cambridge University Press, Cambridge, 1989).
\bibitem{Reimann:91} P.\ Reimann, and P.\ Talkner, Phys.\ Rev.\ E
{\bf 44}, 6348 (1991).
\bibitem{Kaneko:83}
K.\ Kaneko, Prog.\ Theor.\ Phys. {\bf 69}, 403 (1984).
\bibitem{Takesue:84}
S.\ Takesue and K.\ Kaneko, Prog.\ Theor.\ Phys. {\bf 71}, 35 (1984).
\bibitem{Grebogi:87} C.\ Grebogi, E.\ Ott and J.\ Yorke, Physica D {\bf
24}, 243 (1987).
\bibitem{RMP}  R. Badii et al, Rev. Mod. Phys. {\bf 66}, 1389 (1994).
\bibitem{Smel_PC} V. Smelyansky, private communication.


\end{thebibliography}
\end{document}